\documentclass[usenatbib]{mnras}
\usepackage[T1]{fontenc}
\DeclareRobustCommand{\VAN}[3]{#2}
\let\VANthebibliography\thebibliography
\def\thebibliography{\DeclareRobustCommand{\VAN}[3]{##3}\VANthebibliography}
\usepackage{graphicx}	
\usepackage{amsmath}	
\usepackage{amssymb}	
\usepackage{color}
\usepackage{url}
\usepackage{dsfont}
\usepackage{subfig}
\usepackage{CJKutf8}
\usepackage{threeparttable}

\newcommand\lsim{\mathrel{\rlap{\lower4pt\hbox{\hskip1pt$\sim$}}
        \raise1pt\hbox{$<$}}}
\newcommand\gsim{\mathrel{\rlap{\lower4pt\hbox{\hskip1pt$\sim$}}
        \raise1pt\hbox{$>$}}}
\newcommand{\lya}{\ifmmode\mathrm{Ly}\alpha\else{}Ly$\alpha$\fi}
\newcommand{\lyb}{\ifmmode\mathrm{Ly}\beta\else{}Ly$\beta$\fi}
\newcommand{\igm}{\ifmmode\mathrm{IGM}\else{}IGM\fi}
\newcommand{\lae}{\ifmmode\mathrm{LAE}\else{}LAE\fi}
\newcommand{\h}{\ifmmode\mathrm{H}\else{}H\fi}
\newcommand{\hi}{\ifmmode\mathrm{H\,{\scriptscriptstyle I}}\else{}H\,{\scriptsize I}\fi}
\newcommand{\hii}{\ifmmode\mathrm{H\,{\scriptscriptstyle II}}\else{}H\,{\scriptsize II}\fi}
\newcommand{\cmb}{\ifmmode\mathrm{CMB}\else{}CMB\fi}
\newcommand{\qso}{\ifmmode\mathrm{QSO}\else{}QSO\fi}
\newcommand{\eor}{\ifmmode\mathrm{EoR}\else{}EoR\fi}
\newcommand{\heii}{\ifmmode\mathrm{He\,{\scriptscriptstyle II}}\else{}He\,{\scriptsize II}\fi}
\newcommand{\heiii}{\ifmmode\mathrm{He\,{\scriptscriptstyle III}}\else{}He\,{\scriptsize III}\fi}
\newcommand{\ciii}{\ifmmode\mathrm{C\,{\scriptscriptstyle III]}}\else{}C\,{\scriptsize III]}\fi}
\newcommand{\oiii}{\ifmmode\mathrm{O\,{\scriptscriptstyle III}}\else{}O\,{\scriptsize III}\fi}
\newcommand{\aliii}{\ifmmode\mathrm{Al\,{\scriptscriptstyle III}}\else{}Al\,{\scriptsize III}\fi}
\newcommand{\mgii}{\ifmmode\mathrm{Mg\,{\scriptscriptstyle II}}\else{}Mg\,{\scriptsize II}\fi}
\newcommand{\fe}{\ifmmode\mathrm{Fe}\else{}Fe\fi}
\newcommand{\nv}{\ifmmode\mathrm{N\,{\scriptscriptstyle V}}\else{}N\,{\scriptsize V}\fi}
\newcommand{\niv}{\ifmmode\mathrm{N\,{\scriptscriptstyle IV]}}\else{}N\,{\scriptsize IV]}\fi}
\newcommand{\cii}{\ifmmode\mathrm{C\,{\scriptscriptstyle II}}\else{}C\,{\scriptsize II}\fi}
\newcommand{\civ}{\ifmmode\mathrm{C\,{\scriptscriptstyle IV}}\else{}C\,{\scriptsize IV}\fi}
\newcommand{\siv}{\ifmmode\mathrm{Si\,{\scriptscriptstyle IV}}\else{}Si\,{\scriptsize IV}\fi}
\newcommand{\siii}{\ifmmode\mathrm{Si\,{\scriptscriptstyle II}}\else{}Si\,{\scriptsize II}\fi}
\newcommand{\siiii}{\ifmmode\mathrm{Si\,{\scriptscriptstyle III]}}\else{}Si\,{\scriptsize III]}\fi}
\newcommand{\ovi}{\ifmmode\mathrm{O\,{\scriptscriptstyle VI}}\else{}O\,{\scriptsize VI}\fi}
\newcommand{\sioiv}{\ifmmode\mathrm{Si\,{\scriptscriptstyle IV}\,\plus O\,{\scriptscriptstyle IV]}}\else{}Si\,{\scriptsize IV}\,+O\,{\scriptsize IV]}\fi}

\newcommand{\cmmc}{\textsc{\small 21CMMC}}
\newcommand{\cmfst}{\textsc{\small 21CMFAST}}

\usepackage{newtxtext,newtxmath}

\pdfoutput=1

\title[Impact of HMF on inferred EoR astrophysical parameters]{Exploring the role of the halo mass function for inferring astrophysical parameters during reionisation}

\author[B. Greig et al.]{
Bradley Greig$^{1,2,3}$\thanks{E-mail:~bradley.greig@anu.edu.au}, David Prelogovi\'{c}$^{4}$, Jordan Mirocha$^{5,6}$, Yuxiang Qin$^{2,3}$, Yuan-Sen Ting (丁源森)$^{1,7,8,9,10}$ \newauthor \& Andrei Mesinger$^{4}$ \\
$^1$Research School of Astronomy \& Astrophysics, Australian National University, Canberra, ACT 2611, Australia \\
$^2$School of Physics, University of Melbourne, Parkville, VIC 3010, Australia \\
$^3$ARC Centre of Excellence for All-Sky Astrophysics in 3 Dimensions (ASTRO 3D) \\
$^4$Scuola Normale Superiore, Piazza dei Cavalieri 7, I-56125 Pisa, Italy \\
$^5$Jet Propulsion Laboratory, California Institute of Technology, 4800 Oak Grove Drive, Pasadena, CA 91109, USA \\ 
$^6$California Institute of Technology, 1200 E. California Boulevard, Pasadena, CA 91125, USA \\
$^7$School of Computing, Australian National University, Acton ACT 2601, Australia \\
$^8$Department of Astronomy, The Ohio State University, Columbus, OH 45701, USA \\
$^9$Center for Cosmology and AstroParticle Physics (CCAPP), The Ohio State University, Columbus, OH 43210, USA \\
$^{10}$Department of Physics, Faculty of Science, Universiti Malaya, 50603 Kuala Lumpur, Malaysia \\
}

\pubyear{2024}

\begin{document}
\label{firstpage}
\pagerange{\pageref{firstpage}--\pageref{lastpage}}
\begin{CJK}{UTF8}{gkai} 
\maketitle
\end{CJK}

\begin{abstract}
The detection of the 21-cm signal at $z\gtrsim6$ will reveal insights into the properties of the first galaxies responsible for driving reionisation. To extract this information, we perform parameter inference which requires embedding 3D simulations of the 21-cm signal within a Bayesian inference pipeline. Presently, when performing inference we must choose which sources of uncertainty to sample and which to hold fixed. Since the astrophysics of galaxies are much more uncertain than those of the underlying halo-mass function (HMF), we usually parameterise and model the former while fixing the latter. However, in doing so we may bias our inference of the properties of these first galaxies. In this work, we explore the consequences of assuming an incorrect choice of HMF and quantify the relative biases in our inferred astrophysical model parameters when considering the wrong HMF. We then relax this assumption by constructing a generalised five parameter model for the HMF and simultaneously recover these parameters along with our underlying astrophysical model. For this analysis, we use \cmfst{} and perform Simulation-Based Inference by applying marginal neural ratio estimation to learn the likelihood-to-evidence ratio using \textsc{Swyft}. Using a mock 1000 hour observation of the 21-cm power spectrum from the forthcoming Square Kilometre Array, conservatively assuming foreground wedge avoidance, we find assuming the incorrect HMF can bias the recovered astrophysical parameters by up to $\sim3-4\sigma$ even when including independent information from observed luminosity functions. When considering our generalised HMF model, we recover constraints on our astrophysical parameters with a factor of $\sim2-4$ larger marginalised uncertainties. Importantly, these constraints are unbiased, agnostic to the underlying HMF and therefore more conservative.
\end{abstract}

\begin{keywords}
cosmology: theory -- dark ages, reionisation, first stars -- diffuse radiation -- early Universe -- galaxies: high-redshift -- intergalactic medium
\end{keywords}



\section{Introduction}

The formation of the first stars and galaxies, referred to as the cosmic dawn (CD), signifies the end of the cosmic dark ages. The intergalactic medium (IGM), enshrouded by a neutral hydrogen fog following recombination renders the early Universe opaque to most forms of radiation. These first stars and galaxies, emitting copious amounts of ultra-violet (UV) photons, gradually eat away at this fog, leading to a complex morphology of ionised regions embedded in a neutral medium. Over time, as galaxies grow and become more abundant their combined UV photon output almost completely removes this neutral hydrogen fog. This process, is referred to as the Epoch of Reionisation (EoR).

Peering through this fog to directly explore the first generation of stars and galaxies is extremely difficult. Thankfully, we can indirectly infer the properties of the entire population of these first sources via their role in driving the EoR. This is achieved by measuring the temperature and spatial distribution of the neutral hydrogen in the IGM as it is first heated and then gradually disappears due to ionisations over cosmic time. We detect the neutral hydrogen via its 21-cm hyperfine spin-flip transition, which is measured relative to a background radiation source, such as the Cosmic Microwave Background \citep[see e.g.][]{Gnedin:1997p4494,Madau:1997p4479,Shaver:1999p4549,Tozzi:2000p4510,Gnedin:2004p4481,Furlanetto:2006p209,Morales:2010p1274,Pritchard:2012p2958}. Observing this 21-cm signal as a function of time (frequency) reveals a three-dimensional, time-evolving picture of the ionisation and thermal state of the IGM during the early Universe.

Maximising the wealth of astrophysical information in the 21-cm signal requires measuring this spatially varying signal, only achievable using large-scale radio interferometers. In recent years, several of these experiments have begun to report upper-limits on the 21-cm signal via the power spectrum (PS; \citealt{Mertens:2020,Trott:2020,Abdurashidova:2022,HERA:2023}). In turn, these upper-limits have been scrutinised to begin to rule out credible regions of parameter space describing the galaxies responsible for reionisation \citep[e.g.][]{Ghara:2020,Greig:2020,Mondal:2020,Ghara:2021,Greig:2021,Abdurashidova:2022b,HERA:2023}. As we approach the first ever detection of the 21-cm signal it is imperative that our theoretical frameworks and model assumptions are well characterised and understood.

For example, presently when inferring the properties of the galaxies that drive reionisation we typically assume a fixed halo-mass function (HMF)\footnote{We also typically assume fixed cosmological parameters, which if allowed to vary will also have an impact on the inferred astrophysical parameters \citep{Kern:2017}.}. This choice is motivated by the fact that the uncertainties in the astrophysical properties describing the galaxies responsible for reionisation are much larger than the uncertainties in the underlying halo-mass function (HMF). Consequently, by fixing the HMF we can then minimise the computational burden of our inference pipelines. However, as a consequence our inferences on the astrophysical properties describing these galaxies may be biased. In this work, we will explore how biased our astrophysical inferences can be due to this assumption of a fixed HMF by varying several theoretical models with incorrectly assumed HMFs relative to a fiducial mock observation with known HMF. Of course, in practise we do not know the true HMF describing the Universe.. \citet{Mirocha:2021} explored the role of several of these theoretical modelling uncertainties, including the impact of the HMF, on inferring the galaxy properties using the 21-cm global signal. Along with the stellar population synthesis model, the choice of assumed HMF was observed to have a significant impact on the inferred astrophysics. However, the numerical modelling of the 21-cm signal in \citet{Mirocha:2021} employed a simple two-zone IGM model and forgoes modelling the complex 3D nature of the cosmic signal. Therefore, we revisit this, exploring the impact of assuming a fixed HMF in the context of observations of the 21-cm PS, which requires modelling the full 3D signal.

Traditionally, to tackle this Bayesian inference problem we would employ a tool such as \cmmc{} \citep{Greig:2015p3675,Greig:2017p8496,Greig:2018,Park:2019} which performs on-the-fly 3D semi-numerical reionisation simulations using \cmfst{} \citep{Mesinger:2007p122,Mesinger:2011p1123,Murray:2020} within a Monte-Carlo Markov-Chain (MCMC) framework. However, such an approach can be computationally intensive, especially as the number of free parameters increase or in this instance where we are interested in varying the HMF requiring a new MCMC per HMF. Therefore, in this work we utilise  likelihood-free or simulation based inference (SBI; see e.g. \citealt{Cranmer:2020} for a recent review) which adopts machine learning principles to either bypass or model the general likelihood function (avoiding the typically Gaussian assumptions to be able to derive an analytic functional form of the likelihood; see e.g. \citealt{Prelogovic:2023} for detailed discussions). SBI directly connects the forward modelled simulated data to the posterior distribution of the input model parameters or to first `learn' the likelihood from the same forward modelled data which can then be used within an MCMC framework to obtain the model posteriors. There are several benefits of applying these approaches including the removal of the need for an assumed analytic likelihood which relaxes many assumptions and in the process it enables complex (non-Gaussian) summary statistics to be more readily accessed in a Bayesian framework. Further, the forward modelled simulated datasets can be more efficient at sampling the prior volume (removing redundant model evaluations that occur in an MCMC) leading to fewer simulated models to achieve comparable or better performance. In recent years, these approaches have gained in popularity for tackling astrophysical inference from the cosmic 21-cm signal \citep[e.g.][]{Zhao:2022,Zhao:2022b,Prelogovic:2023,Saxena:2023}. 

In this work, we choose to follow the approach of \citet{Saxena:2023} and perform our parameter inference using Marginal Neutral Ratio Estimation (MNRE; \citealt{Miller:2021}) using the publicly available \textsc{\small Python} package, \textsc{\small Swyft}\footnote{https://github.com/undark-lab/swyft} \citep{Miller:2022}. Rather than learning the likelihood, in MNRE the goal is instead to learn the marginal likelihood-to-evidence ratios for each individual parameter combination through its own unique neural network. This approach can be more efficient than sampling the full posterior using an MCMC \citep{Saxena:2023}. Owing to these benefits, this approach has already been applied for several cosmological applications including the CMB \citep{Cole:2022}, gravitational lensing \citep{Montel:2023,Coogan:2024}, gravitational waves \citep{Gagnon:2023,Bhardwaj:2023}, Milky Way stellar streams \citep{Alvey:2023} and supernovae cosmology \citep{Karchev:2023}.

Specifically, for our work we shall utilise \textsc{\small Swyft} to perform astrophysical parameter inference from a mock observation of the 21-cm PS from the forthcoming Square Kilometre Array (SKA; \citealt{Mellema:2013p2975,Koopmans:2015}). For this, we will utilise \cmfst{}s flexible galaxy parameterisation \citep{Park:2019} to make use of UV luminosity functions as additional observational priors. We shall perform several inferences, varying the underlying HMF in order to quantify the bias in the inferred parameters as a result of our assumptions. Following this, making use of \textsc{\small Swyft}s efficiency, we then look to relax our assumption of a fixed HMF by jointly recovering our astrophysical model parameters and those of a generalised HMF functional form.

The remainder of this paper is organised as follows. In Section~\ref{sec:21cm} we summarise our 21-cm simulations using \cmfst{} and provide summary statistics of the 21-cm signal in order to highlight its sensitivity to the underlying choice of HMF. In Section~\ref{sec:setup} we describe our approach for constructing our noisy, simulated data for performing SBI with \textsc{Swyft}. In Section~\ref{sec:fixed} we explore the impact of assuming a fixed HMF on the inferred astrophysical model parameters before considering a generalised function form for the HMF in Section~\ref{sec:general} and simultaneously recovering our astrophysical and HMF model parameters. Finally, in Section~\ref{sec:conclusion} we provide our closing remarks. Unless stated otherwise, all quantities are in co-moving units and we adopt the cosmological parameters:  ($\Omega_\Lambda$, $\Omega_{\rm M}$, $\Omega_b$, $n$, $\sigma_8$, $H_0$) = (0.69, 0.31, 0.048, 0.97, 0.81, 68 km s$^{-1}$ Mpc$^{-1}$), consistent with recent results from the Planck mission \citep{Planck:2020}.

\section{Simulating the 21-cm signal} \label{sec:21cm}

For simulating the 3D 21-cm signal during reionisation we use the semi-numerical simulation code \cmfst{}\footnote{https://github.com/21cmfast/21cmFAST}\citep{Mesinger:2007p122,Mesinger:2011p1123}, specifically the latest public release, v3 \citep{Murray:2020}. In particular, we adopt the flexible galaxy parameterisation of \citet{Park:2019} which describes the UV and X-ray properties of the galaxies responsible for reionisation. Below, we outline the key concepts for simulating the 21-cm signal, with particular focus on the astrophysical parameters that we seek to recover in our parameter inference pipeline and their dependence on the HMF. For more detailed discussions please refer to these aforementioned publications.

\subsection{Galaxy UV properties}

The stellar mass, $M_{\ast}$, of a galaxy is assumed to be directly proportional to its host halo mass, $M_{\rm h}$ \citep[e.g.][]{Kuhlen:2012p1506,Dayal:2014b,Behroozi:2015p1,Mitra:2015,Mutch:2016,Ocvirk:2016,Sun:2016p8225,Yue:2016,Hutter:2020},
\begin{eqnarray} \label{}
M_{\ast}(M_{\rm h}) = f_{\ast}\left(\frac{\Omega_{\rm b}}{\Omega_{\rm m}}\right)M_{\rm h},
\end{eqnarray}
with the fraction of galactic gas in stars, $f_{\ast}$, also dependent on the host halo mass via a power-law\footnote{This follows directly from the mean behaviour of $M_{\ast}$ and $M_{\rm h}$ obtained from both semi-empirical fits to observations \citep[e.g.][]{Harikane:2016,Tacchella:2018,Behroozi:2019,Stefanon:2021} and semi-analytic model predictions \citep[e.g][]{Mutch:2016,Yung:2019,Hutter:2020}.} characterised by two free model parameters, the index, $\alpha_{\ast}$, and the expression is normalised via $f_{\ast, 10}$ for a host dark matter halo mass of $10^{10}$~$M_{\odot}$,
\begin{eqnarray} \label{}
f_{\ast} = f_{\ast, 10}\left(\frac{M_{\rm h}}{10^{10}\,M_{\odot}}\right)^{\alpha_{\ast}}.
\end{eqnarray}
Dividing this stellar mass by a characteristic time-scale, $t_{\ast}$, then yields the star-formation rate (SFR) for these galaxies,
\begin{eqnarray} \label{eq:sfr}
\dot{M}_{\ast}(M_{\rm h},z) = \frac{M_{\ast}}{t_{\ast}H^{-1}(z)},
\end{eqnarray}
where $H^{-1}(z)$ is the Hubble time and $t_{\ast}$ $\in[0.05,1]$.

The fraction of UV photons escaping into the IGM, $f_{\rm esc}$, is also assumed to be governed by a power-law relation with halo mass,
\begin{eqnarray} \label{}
f_{\rm esc} = f_{\rm esc, 10}\left(\frac{M_{\rm h}}{10^{10}\,M_{\odot}}\right)^{\alpha_{\rm esc}},
\end{eqnarray}
yielding an additional two parameter, index, $\alpha_{\rm esc}$ and normalisation, $f_{\rm esc, 10}$.

Importantly, not all haloes are capable of contributing to reionisation owing to either internal feedback and/or inefficient cooling that can stymie star-formation in low mass haloes. To mimic this behaviour, an effective duty-cycle is adopted,
\begin{eqnarray} \label{eq:duty}
f_{\rm duty} = {\rm exp}\left(-\frac{M_{\rm turn}}{M_{\rm h}}\right).
\end{eqnarray}
where $(1 - f_{\rm duty})$ corresponds to the fraction of suppressed star-forming galaxies below a characteristic mass scale $M_{\rm turn}$ \citep[e.g.][]{Shapiro:1994,Giroux:1994,Hui:1997,Barkana:2001p1634,Springel:2003p2176,Mesinger:2008,Okamoto:2008p2183,Sobacchi:2013p2189,Sobacchi:2013p2190}.

The primary impact of our assumed HMF is through the production rate of UV photons, which directly feeds into the corresponding UV luminosity functions (LFs). The above galaxy parameterisation connects to UV LFs by first computing the non-ionising UV LFs;
\begin{eqnarray} \label{eq:UVLF}
\phi(M_{\rm UV}) = \left[f_{\rm duty}\frac{{\rm d}n}{{\rm d}M_{\rm h}}\right] \left | \frac{{\rm d}M_{\rm h}}{{\rm d}M_{\rm UV}} \right |.
\end{eqnarray}
The quantity in square brackets corresponds to the number density of active star-forming galaxies and is directly proportional to the HMF. Therefore, including the UV LFs when exploring the role of assuming a fixed HMF will be important. The final term performs the conversion between halo mass and UV magnitude. Our defined star-formation rate (Equation~\ref{eq:sfr}) is then related to the UV luminosity ($L_{\rm UV}$) via:
\begin{eqnarray}
\dot{M}_{\ast}(M_{\rm h},z) = \mathcal{K}_{\rm UV} \times L_{\rm UV},
\end{eqnarray}
where $\mathcal{K}_{\rm UV} = 1.15\times10^{-28}\,M_{\odot}\,{\rm yr}^{-1}/{\rm erg\,s^{-1}}\,{\rm Hz}^{-1}$ is a conversion factor \citep{Madau:2014} and the UV luminosity is related to AB magnitude via the standard relation \citep{Oke:1983}:
\begin{eqnarray}
{\rm log}_{10}\left(\frac{L_{\rm UV}}{{\rm erg\,s^{-1}\,Hz^{-1}}} \right) = 0.4 \times (51.63 - M_{\rm UV}).
\end{eqnarray}

In summary, this model contains six free parameters to describe the UV properties of the galaxies, namely, $f_{\ast, 10}$, $\alpha_{\ast}$, $f_{\rm esc, 10}$, $\alpha_{\rm esc}$, $t_{\rm ast}$ and $M_{\rm turn}$.

\subsection{Galaxy X-ray properties}

In addition to UV photons, these first galaxies emit X-ray photons originating from stellar remnants from earlier episodes of star-formation. These escaping X-ray photons, with their long mean-free paths, are capable of traversing large distances and heating the adiabatically cooling IGM. Within \cmfst{}, X-ray heating is modelled by first computing the angle-averaged specific X-ray intensity, $J(\boldsymbol{x}, E, z)$, (in erg s$^{-1}$ keV$^{-1}$ cm$^{-2}$ sr$^{-1}$) within each simulation voxel,
\begin{equation} \label{eq:Jave}
J(\boldsymbol{x}, E, z) = \frac{(1+z)^3}{4\pi} \int_{z}^{\infty} dz' \frac{c dt}{dz'} \epsilon_{\rm X}  e^{-\tau}.
\end{equation}
This calculates the contribution from the co-moving X-ray specific emissivity, $\epsilon_{\rm X}(\boldsymbol{x}, E_e, z')$, emitted at earlier times, $E_{\rm e} = E(1 + z')/(1 + z)$, while also accounting for their attenuation by the IGM, $e^{-\tau}$, as they traverse the vast distances through the Universe. The specific emitted emissivity is then,
\begin{equation} \label{eq:emissivity}
\epsilon_{\rm X}(\boldsymbol{x}, E_{\rm e}, z') = \frac{L_{\rm X}}{\rm SFR} \left[ (1+\bar{\delta}_{\rm nl}) \int^{\infty}_{0}{\rm d}M_{\rm h} \frac{{\rm d}n}{{\rm d}M_{\rm h}}f_{\rm duty} \dot{M}_{\ast}\right],
\end{equation}
where $\bar{\delta}_{\rm nl}$ is the mean, non-linear overdensity in a shell centred on the simulation cell $(\boldsymbol{x}, z)$, $L_{\rm X}/{\rm SFR}$ (erg s$^{-1}$ keV$^{-1}$ $M^{-1}_{\odot}$ yr) is the specific X-ray luminosity per unit star formation escaping the host galaxies and the quantity in square brackets is the SFR density along the light-cone. This latter quantity depends on our assumed choice for the halo mass function (HMF) highlighting that our modelled X-ray emissivity is directly impacted by our HMF choice. Either a change to the HMF could be compensated for by a corresponding change to $L_{\rm X}/{\rm SFR}$ or through the stellar properties of the UV galaxies, namely $M_{\rm turn}$ (through $f_{\rm duty}$), $f_{\ast, 10}$, $\alpha_{\ast}$ or $t_{\ast}$.

$L_{\rm X}/{\rm SFR}$ acts as the normalisation of the X-ray emissivity and is modelled with a power-law dependence with photon energy, $L_{\rm X} \propto E^{- \alpha_X}$, with $\alpha_{\rm X}$ denoting the index of the spectral energy distribution describing the source of X-rays. In this work, high-mass X-ray binaries are assumed to dominant, thus we adopt $\alpha_{\rm X} = 1$ consistent with local Universe observations \citep{Mineo:2012p6282,Fragos:2013p6529,Pacucci:2014p4323}. Finally, this specific X-ray luminosity is then normalised by the integrated soft-band ($<2$~keV) luminosity per SFR (in erg s$^{-1}$ $M^{-1}_{\odot}$ yr),
\begin{equation} \label{eq:normL}
  L_{{\rm X}<2\,{\rm keV}}/{\rm SFR} = \int^{2\,{\rm keV}}_{E_{0}} dE_e ~ L_{\rm X}/{\rm SFR} ~,
\end{equation}
where $E_0$ corresponds to the energy threshold below which X-ray photons are absorbed by the host galaxy. Both $ L_{{\rm X}<2\,{\rm keV}}/{\rm SFR}$ and $E_0$ are free parameters in our astrophysical model.

\subsection{Ionisation and Thermal State of the IGM}

Computing the thermal and ionisation states of the IGM requires the velocity and evolved density fields which are determined using second-order Lagrange perturbation theory \citep{Scoccimarro:1998p7939}. First, the thermal state of the neutral IGM is calculated by determining the IGM spin temperature, $T_{S}$, which is obtained by self-consistently computing X-ray heating and ionisations along with heating and cooling due to structure formation, Compton scattering by CMB photons and heating due to partial ionisations. $T_{S}$ is set by the weighted mean,
\begin{eqnarray}
T^{-1}_{\rm S} = \frac{T^{-1}_{\rm CMB} + x_{\alpha}T^{-1}_{\alpha} + x_{\rm c}T^{-1}_{\rm K}}{1 + x_{\alpha} + x_{\rm c}},
\end{eqnarray}
where $T_{\rm K}$, $T_{\alpha}$ and $T_{\rm CMB}$ are the gas, Lyman-$\alpha$ (Ly$\alpha$) colour and CMB temperatures, respectively, and $T_{S}$ depends both on the local gas density and the \lya{} radiation intensity. The \lya{} background is the cumulative sum of X-ray excitations of neutral hydrogen atoms and the direct stellar emission of Lyman band photons by the first sources with $x_{\alpha}$ denoting the Wouthuysen-Field coupling coefficient \citep{Wouthuysen:1952p4321,Field:1958p1} and $x_{\rm c}$ the collisional coupling coefficient between the free electrons and CMB photons.

The ionisation field is then determined by applying excursion-set theory \citep{Furlanetto:2004p123} on the evolved density field which compares the cumulative number of ionising photons, $n_{\rm ion}$, to the total number of neutral hydrogen atoms plus cumulative recombinations, $\bar{n}_{\rm rec}$ \citep{Sobacchi:2014p1157} within spheres of decreasing radii, $R$, and corresponding overdensity, $\delta_{R}$.  An individual simulation voxel is then considered ionised if
\begin{eqnarray} \label{eq:ioncrit}
\bar{n}_{\rm ion}(\boldsymbol{x}, z | R, \delta_{R}) \geq (1 + \bar{n}_{\rm rec})(1-\bar{x}_{e}),
\end{eqnarray}
where $(1-\bar{x}_{e})$ accounts for secondary ionisations by X-ray photons. The cumulative number of ionisations is determined by,
\begin{eqnarray} \label{eq:ioncrit2}
n_{\rm ion} = \bar{\rho}^{-1}_b\int^{\infty}_{0}{\rm d}M_{\rm h} \frac{{\rm d}n(M_{h}, z | R, \delta_{R})}{{\rm d}M_{\rm h}}f_{\rm duty} \dot{M}_{\ast}f_{\rm esc}N_{\gamma/b},
\end{eqnarray}
where $\bar{\rho}_b$ is the mean baryon density and the total number of ionising photons per stellar baryon is given by $N_{\gamma/b}$ which is assumed to be $N_{\gamma/b}=5000$ consistent with a Salpeter initial mass function \citep{Salpeter:1955}. As one would expect, $n_{\rm ion}$ is directly connected to our adopted HMF. Therefore, the relative amplitude and slope of the HMF (i.e. number density of haloes of a given mass) will have a notable impact on the total number of ionisations and thus primarily the ionisation state of the IGM. Consequently, this implies any of the six free model parameters describing the UV galaxies could be impacted by the assumed choice of HMF.

\subsection{21-cm Brightness Temperature}

The intensity of the observed 21-cm radiation, or brightness temperature, $\delta T_{\rm b}(\nu)$, is set by the optical depth, $\tau_{\nu_{0}}$, of the neutral hydrogen gas (where $\nu_{0}$ is the frequency of the 21-cm signal) and the background radiation passing through it (e.g. CMB; \citealt{Furlanetto:2006p209});
\begin{eqnarray} \label{eq:21cmTb}
\delta T_{\rm b}(\nu) &=& \frac{T_{\rm S} - T_{\rm CMB}(z)}{1+z}\left(1 - {\rm e}^{-\tau_{\nu_{0}}}\right)~{\rm mK},
\end{eqnarray}
and
\begin{eqnarray}
\tau_{\nu_{0}} &\propto& (1+\delta_{\rm nl})(1+z)^{3/2}\frac{x_{\hi{}}}{T_{\rm S}}\left(\frac{H}{{\rm d}v_{\rm r}/{\rm d}r+H}\right).
\end{eqnarray}
The optical depth depends on the neutral hydrogen fraction, $x_{\hi{}}$, local gas overdensity, $\delta_{\rm nl} \equiv \rho/\bar{\rho} - 1$, the Hubble parameter, $H(z)$, and the line-of-sight gradient of the peculiar velocity. This quantity is evaluated at the redshift $z = \nu_{0}/\nu - 1$ and for ease of notation we have dropped the explicit spatial dependence of the quantities.

\subsection{The halo mass function (HMF)}

As outlined previously, the timing and bias of the sources producing UV and X-ray photons is degenerate with our modelled astrophysical parameters which is in turn degenerate with the HMF. In \cmfst{} the HMF is explicitly calculated using an analytic expression, therefore it is straightforward to interchange for various analytic HMFs in the literature. By default, \cmfst{} assumes the \citet[][hereafter ST]{Sheth:2001} ellipsoidal collapse model to compute the number density of haloes. Throughout this work, we adopt ST as our default HMF. For completeness of the purely analytically derived HMFs we shall also consider the \citet[][hereafter PS]{Press:1974} HMF.

Following these, within the literature there have been many analytic fitting functions created to match outputs of $N$-body simulations \citep[e.g.][]{Jenkins:2001,Warren:2006,Reed:2007,Tinker:2008,Tinker:2010,Angulo:2012,Watson:2013,Diemer:2020}. These can vary considerably in their functional forms with differences arising from the fundamentals of the $N$-body simulations (e.g. gravity solver and/or definition of what constitutes a halo along with the halo finder itself). Importantly, the vast majority of these $N$-body simulations are not designed for high-$z$ studies, which is not unsurprising given the wealth of observations at low-$z$ (and lack thereof at high-$z$) from which to calibrate the $N$-body results against\footnote{In fact, even at low redshifts the HMFs are known to differ non-negligibly \citep{Murray:2013}.}.Therefore, the extrapolation of these HMFs out to the high-$z$'s necessary for studying reionisation can lead to significant differences in the number density of sources which can have a significant impact on the inferred astrophysical parameters \citep{Mirocha:2021}.

In addition to the ST and PS HMFs, we also consider the analytic fitting functions of \citet[][hereafter Tinker]{Tinker:2010}, \citet[][hereafter Angulo]{Angulo:2012} and the redshift evolving \citet[][hereafter Watson-$z$]{Watson:2013}. The choice is somewhat arbitrary, but is designed to broadly cover the breadth of variation in the 21-cm signal during reionisation. For example, we found the \citet{Diemer:2020} model to be very similar to that of Watson-$z$ while equally the Angulo and \citet{Jenkins:2001} models were also very similar. 

\begin{figure*}
	\includegraphics[trim = 0.05cm 0cm 0cm 0.5cm, scale = 0.59]{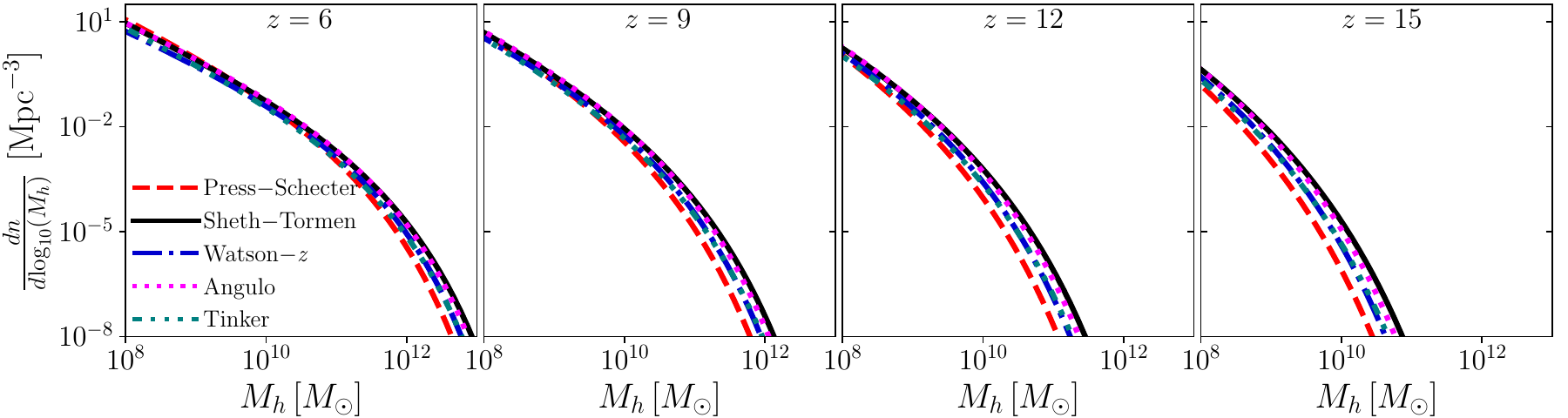}
	\includegraphics[trim = 0.05cm 0.3cm 0cm 0cm, scale = 0.59]{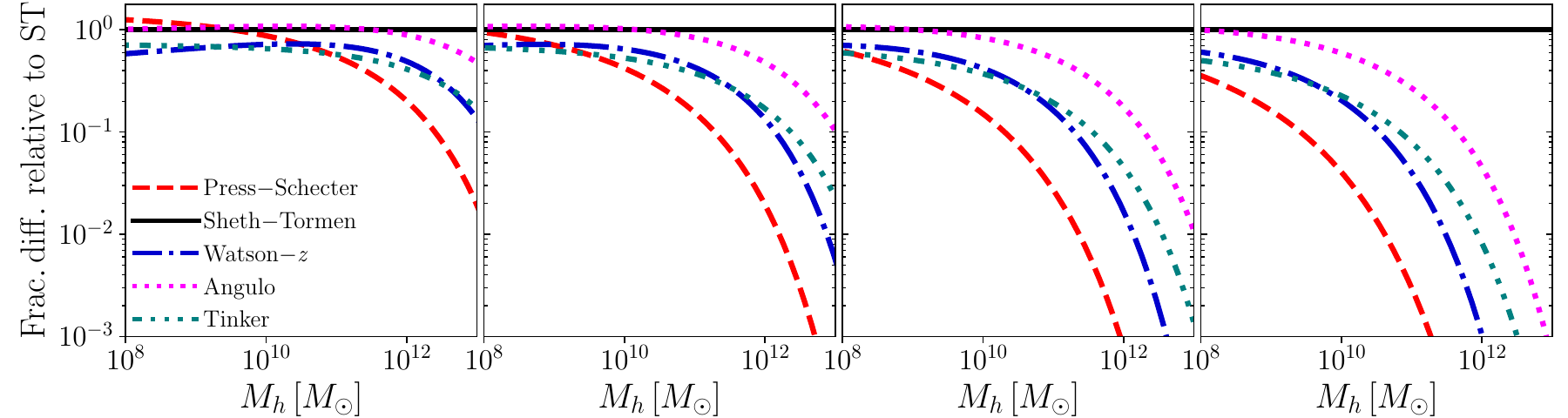}
    \caption{The halo mass functions at $z=6,9,12$ and~15 for the five different models we consider in this work. In particular, we consider the Sheth-Tormen (black solid), Press-Schecter (red dashed), redshift evolving Watson (blue dot-dashed), Angulo (magenta dotted) and the Tinker (teal triple-dot-dashed) HMFs. In the bottom row, we provide the fractional difference between the various HMFs and that of the ST HMF.}
    \label{fig:HMF}
\end{figure*}

In Figure~\ref{fig:HMF} we provide the corresponding HMFs for our selected models at $z = $~6, 9, 12 and 15. Additionally, for ease of comparison we also provide the fractional difference between the various HMFs and the ST HMF. At $z=6$ all models are fairly consistent with one another up until halo masses of $10^{11}\,M_{\odot}$ following which the various models begin to diverge for increasing mass. Although these large mass haloes do not significantly contribute to the total number of ionising photons, they contribute to setting $f_{\ast}$ through the UV LFs and thus small differences in the HMF on these mass scales can still bleed into our inference of the astrophysical parameters. As noted earlier, as we extrapolate these models out to higher redshifts, the relative differences are amplified and extend throughout the entire mass range. At $z=9$ the HMFs are only consistent below $10^{9.5}\,M_{\odot}$ whereas by $z=15$ significant differences occur over the entire mass range. Further, the slope of the different models also differs, resulting in different abundances of sources within different mass bins. Combined, these differences will have a sizeable impact on the 21-cm signal.

\subsection{Mock 21-cm Observation}

\begin{table*}
 \caption{The astrophysical model parameters for our mock 21-cm observation, assuming a \citet[][]{Sheth:2001} HMF, based on the constraints from \citet{Qin:2021}.}
 \label{tab:mock}
\begin{tabular}{ccccccccc}
  \hline
 & ${\rm log}_{10}(f_{\ast,10})$ & $\alpha_{\ast}$ & ${\rm log}_{10}(f_{{\rm esc},10})$ & $\alpha_{\rm esc}$ & ${\rm log}_{10}(M_{\rm turn})$ & $t_{\ast}$ & ${\rm log}_{10}\frac{L_{X<2\,{\rm keV}}}{\rm SFR}$ & $E_{0}$\\
 & & & & & ($M_{\odot}$) & & (erg s$^{-1}$ $M^{-1}_{\odot}$ yr$^{-1}$) & (keV) \\
  \hline
Mock Observation & -1.1 & 0.5 & -1.30 & -0.35 & 8.55 & 0.5 & 40.5 & 0.5 \\
Prior ranges & [-3.0,0.0] & [-0.5, 1.0] & [-3.0,0.0] & [-1.0, 0.5] & [8.0, 10.0] & [0.05,1.0] & [38.0, 42.0] & [0.1,1.5] \\
  \hline
 \end{tabular}
\end{table*}

In order to quantify the relative impact of our HMF choice on our inferred astrophysical parameters, we must define a mock observation of the 21-cm signal\footnote{Theoretically one could also consider the impact of varying the underlying transfer function used for modelling the matter power spectrum. For example, using \cmfst{} we assume a matter power spectrum defined by \citep{Eisenstein:1999}. However, more accurate transfer functions exist which rely on Boltzmann solvers such as CAMB \citep[][]{Lewis:2000} or CLASS \citep[][]{Blas:2011}. Relatively speaking the differences are more minor, with differences consistent with the those between the ST and Angulo HMFs. Nevertheless, they will contribute marginally to biasing our inferred results.}. For this, we adopt a set of astrophysical parameters that closely resemble the recovered model of \citet{Qin:2021} based on the cumulative distribution functions of the \lya{} forest \citep{Bosman:2018} combined with existing constraints on the electron scattering optical depth, $\tau_{e}$ from the CMB \citep{Planck:2020}, UV LFs (see Section~\ref{sec:UVLFs}) and dark fraction \citep{McGreer:2015p3668}. In Table~\ref{tab:mock} we summarise the corresponding model parameters of our mock observation, along with the adopted prior ranges for each of these parameters based on previous work \citep{Park:2019,Greig:2020}. Note, throughout this work, these astrophysical parameters always remained fixed, and we only ever change the underlying HMF when differentiating models. For this mock observation we adopt a ST HMF to describe the number density of ionising sources. Note, for this section, to ease qualitative discussions, all 21-cm models are generated assuming the same initial conditions. 

\subsubsection{Global signal}

\begin{figure*}
	\includegraphics[trim = 0.3cm 0.6cm 0cm 0.5cm, scale = 0.435]{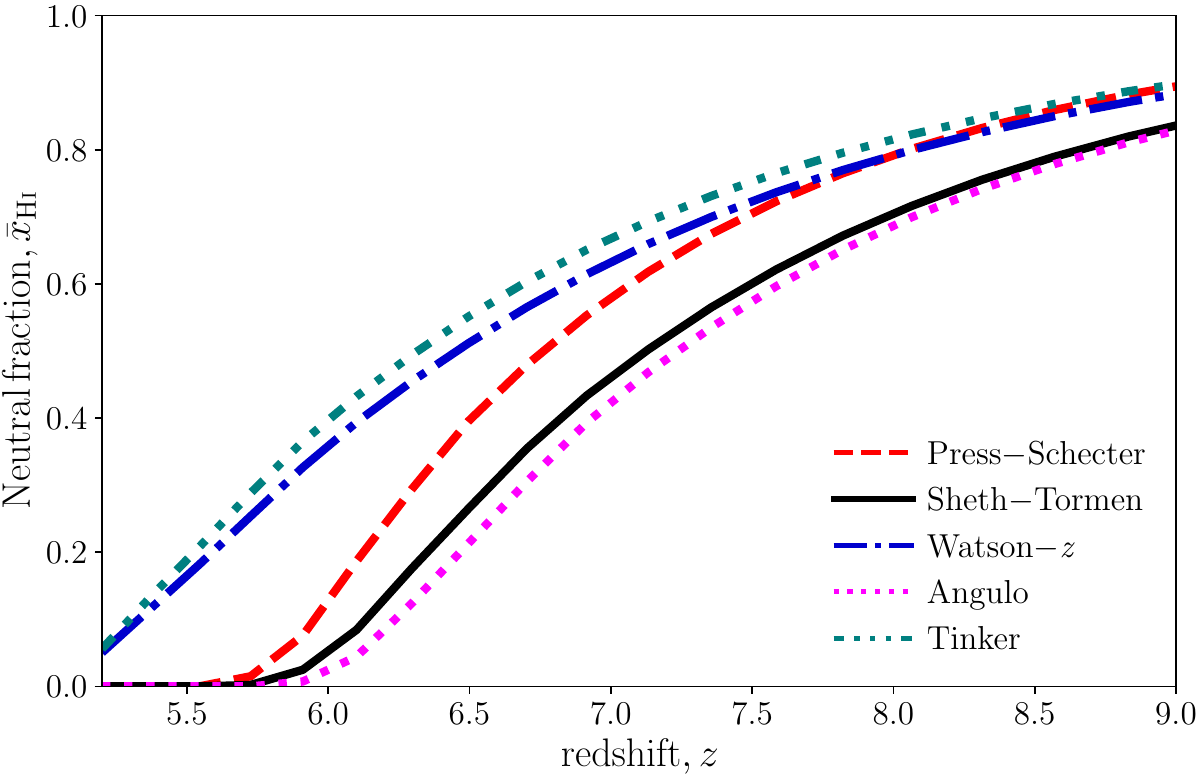}
	\includegraphics[trim = 0.2cm 0.6cm 0cm 0.5cm, scale = 0.435]{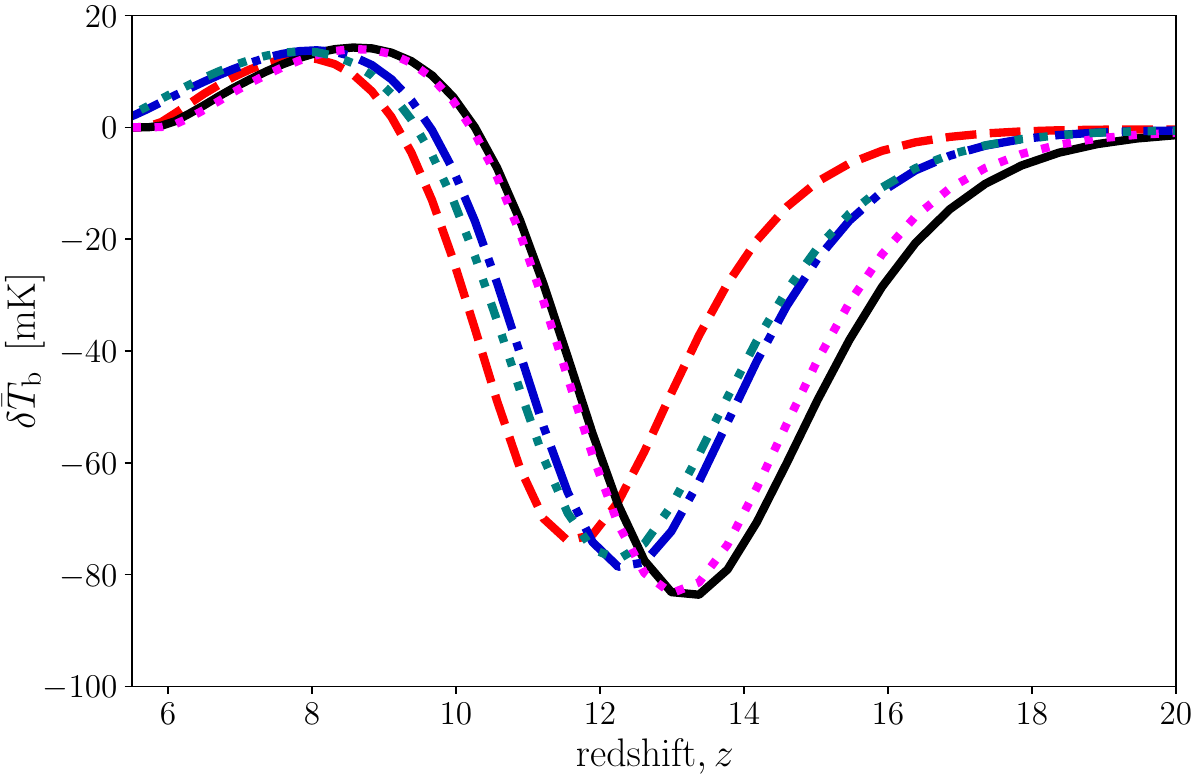}	
    \caption{The global history of reionisation (left panel) and the mean brightness temperature (right panel) for the same astrophysical model parameters (see Table~\ref{tab:mock}) but varying the underlying halo mass function (HMF). In particular, we consider the Sheth-Tormen (black solid; our mock 21-cm observation), Press-Schecter (red dashed), redshift evolving Watson (blue dot-dashed), Angulo (magenta dotted) and the Tinker (teal triple-dot-dashed) HMFs. Note, for this visual comparison, all 21-cm light-cones are generated using the same initial conditions for our fiducial 250 Mpc simulations.}
    \label{fig:history}
\end{figure*}

In Figure~\ref{fig:history} we demonstrate the global history of reionsation (left panel) and the mean brightness temperature signal (right panel) as a function of redshift for our mock 21-cm model assuming a ST HMF (solid black curves). Additionally, we also provide the corresponding curves for our four other HMFs; PS (red, dashed), Watson-$z$ (blue, dot-dashed), Angulo (magenta, dotted) and Tinker (teal, triple-dot-dashed). Immediately evident from this figure are the considerably large differences in the corresponding timing of reionisation along with both the amplitude of the global 21-cm signal and the corresponding redshifts of the various features.

Relative to our fiducial model (ST), the mid-point of reionisation can differ by as much as $\Delta z\sim~1$ depending on the choice of HMF, with both the Tinker and Watson-$z$ models resulting in a delayed and slower reionisation. This is due to the $\sim50$ per cent reduction in the number of sources (e.g. Figure~\ref{fig:HMF}) at the low-mass end. The differences in the HMF choices corresponds to a shift in the electron scattering optical depth of at most $\Delta \tau_{\rm e}\pm0.005$ relative to the fiducial model of $\tau_{\rm e}\sim0.05$. Equally, this difference corresponds to a $\Delta z\sim 1$ shift in the features in the global brightness temperature. The PS model exhibits a notably different reionisation history and 21-cm global signal. At high-$z$, PS considerably under predicts the number of haloes, resulting in the most delayed \lya{} decoupling ($z\sim15$) and absorption trough ($z\sim11$). However, the PS HMF more rapidly evolves on the low-mass end with decreasing redshift, therefore although reionisation starts latest for PS, its duration occurs on the shortest time-scale for any model due to the more rapidly increasing abundance of sources (largest gradient in the reionisation history). It is quite clear from these observations that the various HMFs can produce relatively drastic differences in the globally averaged 21-cm signal.

\subsubsection{21-cm power spectrum}

\begin{figure*}
	\includegraphics[trim = 0.5cm 0.6cm 0cm 0.5cm, scale = 0.7]{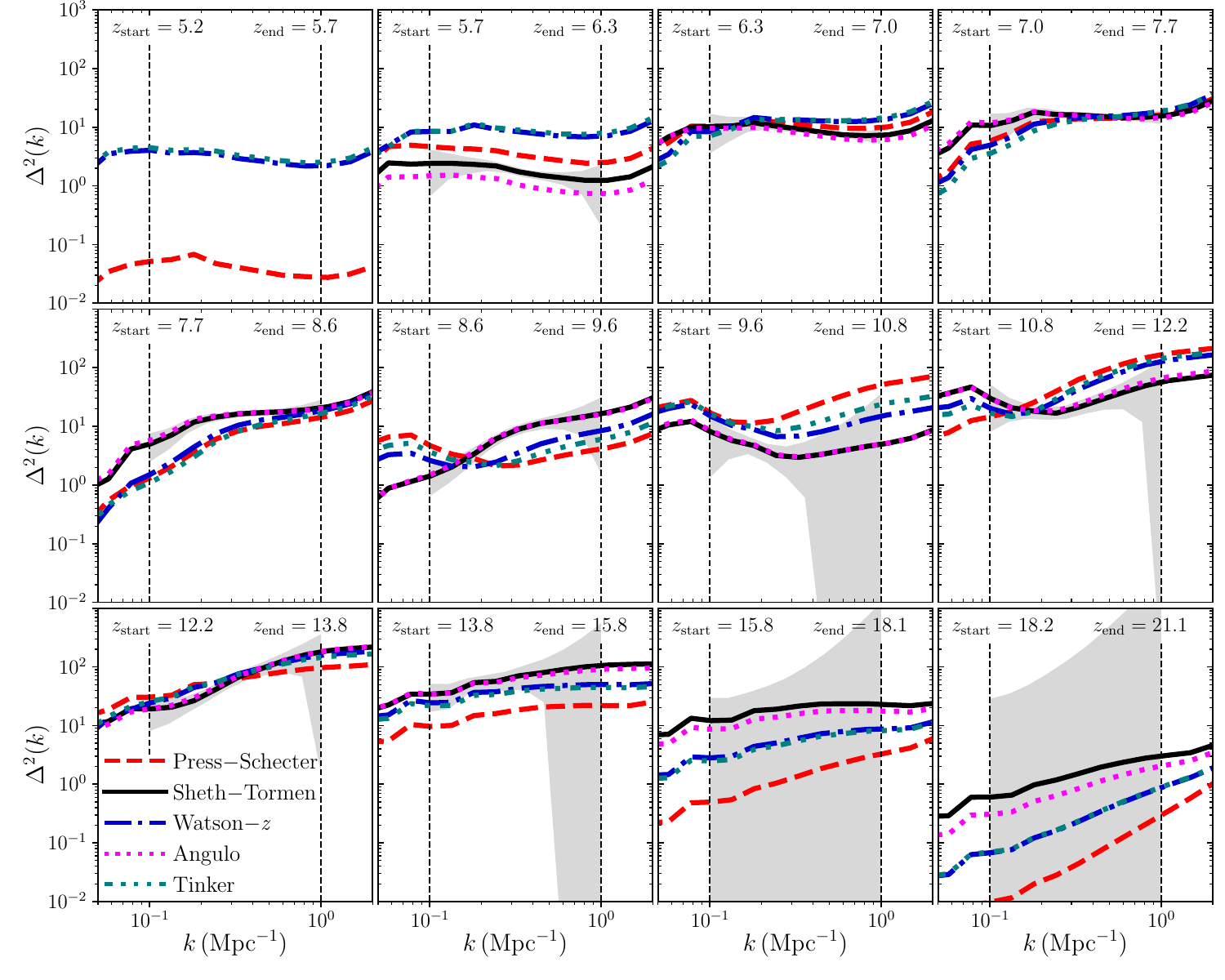}
    \caption{The 21-cm power spectrum (PS) obtained from a 3D 21-cm light-cone generated by \cmfst{} for our fiducial astrophysical parameter set, but varying the underlying HMF. Here, we consider the 21-cm PS obtained from equal comoving length chunks along the 21-cm light-cone. Note, for this we show the 21-cm PS obtained from the full simulation cube, not including any instrumental effect (e.g. foreground wedge) or thermal noise. However, as a visual guide the shaded regions correspond to the 68th percentile uncertainties for a mock 1000 hr observation with the SKA. The vertical dashed lines denote the Fourier scales used within our inference pipeline ($k=0.1$ and $k=1.0$~Mpc$^{-1}$, respectively). Note, for this visual comparison, all 21-cm light-cones are generated using the same initial conditions.}
    \label{fig:21cmPS}
\end{figure*}

In Figure~\ref{fig:21cmPS}, we provide the 21-cm PS for our astrophysical model, again varying the underlying HMF. For this, we take the 3D 21-cm light-cone output by \cmfst{} and break it up into pieces along the line-of-sight, with the start and end redshifts of each cube denoted at the top of each panel. The vertical dashed lines correspond to the spatial scales between which we use the 21-cm PS for our astrophysical parameter inference ($k=0.1$ and $k=1.0$~Mpc$^{-1}$). Finally, the shaded regions denote the 68th percentile uncertainties obtained from a 1000 hr observation with the SKA using \textsc{\small 21cmSense} \citep{Pober:2013p41,Pober:2014p35}. These are provided here purely as a visual aid to provide context to the relative differences in the 21-cm PS due to the choice of HMF. Further, for this figure, we assume perfect foreground removal in order to highlight variations to the 21-cm PS coming directly from the underlying changes to the HMF. In Section~\ref{sec:sim_data} we outline how we model the instrumental effects and include them in our forward-modelled simulations for our parameter inference.

Like previously, for the same fixed astrophysical parameter set the choice of HMF can have a significant impact on the 21-cm PS. Because the choice of HMF impacts the relative timing of the various 21-cm milestones (e.g. reionisation, X-ray heating or \lya{} decoupling), for the same fixed redshift (frequency) we are effectively comparing the 21-cm signal at a different evolutionary stage. For example, in most panels the PS, Watson-$z$ and Tinker models lag behind the mock ST model by almost one entire panel. Relative to the observational uncertainties from the SKA, this can result in differences in excess of several $\sigma$, especially during reionisation (e.g. $z\lesssim10$), when the observational uncertainties are smallest. Even at higher redshifts, when the observational uncertainties are much larger due to increasing thermal noise, certain models can still differ by several $\sigma$ due to the relatively large amplitude signal during the X-ray heating epoch. Therefore, when performing inference and changing the underlying HMF we will be biased to astrophysical parameters that can result in the timing of these cosmic milestones occurring earlier to match that of our fiducial ST model.

Even the Angulo model, which is very similar to the fiducial ST model can differ in excess of the SKA observational uncertainties. For example, during reionisation (e.g. $z\lesssim10$) the Angulo HMF produces slightly more haloes relative to the ST HMF at $\lesssim10^{10}\,M_{\odot}$ (see e.g. Fig~\ref{fig:HMF}). As a result, reionisation begins slightly earlier and proceeds slightly faster (see Figure~\ref{fig:history}). Even by $z<7$ these marginal differences in the HMF can produce 21-cm PS that differ by more than the observational uncertainties. Therefore, the Angulo model will serve as a demonstration of how sensitive our HMF choice is to the inferred astrophysical parameters.

\subsubsection{UV LFs} \label{sec:UVLFs}

Finally, we also consider the impact of our choice of HMF on the UV LFs. In \citet{Park:2019} it was demonstrated that the unique and complimentary information provided by observed high-$z$ UV LFs can improve on the overall constraining power of the underlying model (see also \citealt{Mirocha:2017}). Essentially, this improvement stems from the sensitivity of the UV LFs to the stellar parameters, $f_{\ast}$, breaking the degeneracy with the escape-fraction, $f_{\rm esc}$ when it comes to the UV ionisations. Importantly, since the UV LFs also depend on the HMF, our relative improvements also depend on our assumed HMF.

\begin{figure*}
	\includegraphics[trim = 0.3cm 0.3cm 0cm 0.5cm, scale = 0.595]{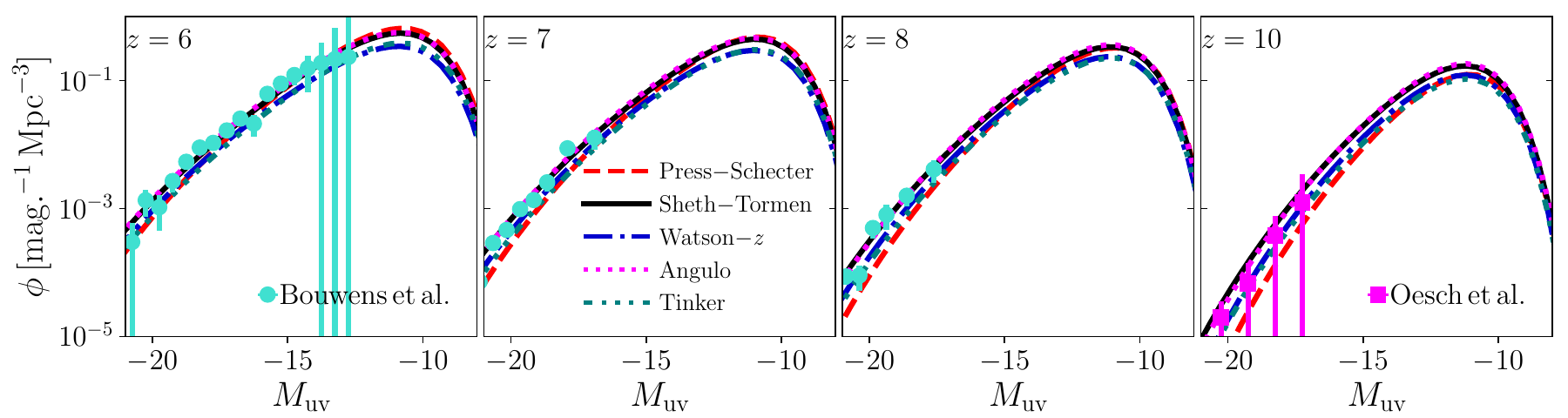}
    \caption{The corresponding UV LFs, assuming the same fixed astrophysical parameter set, but varying the underlying HMF. We show the UV LFs at $z=6,7,8$ and~10 which correspond to the redshifts of existing observational data.}
    \label{fig:LF}
\end{figure*}

In Figure~\ref{fig:LF} we demonstrate the sensitivity of the modelled UV LFs obtained using \cmfst{} for our mock astrophysical parameter set varying only our choice of HMF. For reference, we additionally include a selection of observed UV LFs at $z=6$ \citep{Bouwens:2017}, $z=7$ and 8 \citep{Bouwens:2015} and $z=10$ \citep{Oesch:2018}. Note, this choice follows that of \citet{Park:2019}, based on limiting the systematic differences between the various results in the literature due to how the statistical and observational uncertainties are determined. Ideally, one should combine all results in the literature to obtain a mean UV LF that characterises the scatter across the various observations and pipelines, resulting in a more conservative and unbiased UV LF. However, for now we restrict our results to these existing UV LFs. Additionally, for the purposes of our modelling we ignore the impact of dust on the bright UV galaxies, imposing a limit of $M_{\rm UV} < -20$ below which we argue the galaxies are dust-free \citep[see][for further discussions]{Park:2019}. However, this may not be the case, and any non-zero dust contribution would result in a reddening of the UV slope biasing the inference on $f_{\ast}$. Therefore, in future we aim to explore the inclusion of dust in our model and to make use of the bright-end of the UV LFs. For example, one could envisage utilising either empirical or physically-motivated prescriptions for dust and their connection to the UV properties of the galaxies that have been extensively explored in the literature \citep[e.g.][]{Meurer:1999,Wilkins:2013,Bouwens:2014,Mancini:2016,Zimu:2016,Ma:2019,Qiu:2019,Mirocha:2020b,Zhao:2024}.

For these specific UV LFs, we can see that the variations due to the HMFs are larger than the corresponding statistical uncertainties, therefore the UV LFs have sufficient constraining power to differentiate between the HMF models. Note however, this would be notably reduced if we averaged across all UV LFs and the various systematic uncertainties. For reference, the turn-over in the UV LFs corresponds to our choice for $M_{\rm turn} = 5\times10^{8}\,M_{\odot}$, indicating that the UV LFs are sensitive primarily to the HMF on mass scales greater than this. As highlighted earlier, the extrapolation of the HMFs out to higher redshifts increases the relative differences, which we observe in the case of the UV LFs. Finally, as noted previously changes to the underlying HMF can be compensated by increasing the number of UV photons escaping the galaxies. However, since the UV LFs constrain the stellar contribution through $f_{\ast}$, this implies that when we combine observations of the UV LFs with the 21-cm signal, the largest biases should appear via the escape fraction, $f_{\rm esc}$ as the UV LFs are independent of $f_{\rm esc}$.

\section{Simulation setup} \label{sec:setup}

Having outlined our fiducial astrophysical model and how the HMF impacts the summary statistics of the 21-cm signal, we now focus on discussing our inference pipeline setup.

\subsection{Parameter inference with \textsc{\small Swyft}}

When it comes to parameter inference, our goal is to determine the probability distribution of obtaining our model parameters, $\boldsymbol{\theta}$, given an observation, $\boldsymbol{x}$. This probability distribution, called the posterior, $p(\boldsymbol{\theta}|\,\boldsymbol{x})$, is given by Bayes' theorem,
\begin{eqnarray}
p(\boldsymbol{\theta}|\,\boldsymbol{x}) = \frac{p(\boldsymbol{x}|\,\boldsymbol{\theta})}{p(\boldsymbol{x})}p(\boldsymbol{\theta})
\end{eqnarray}
where $p(\boldsymbol{x}|\,\boldsymbol{\theta})$ denotes the likelihood to obtain the observed measurement $\boldsymbol{x}$ given the model parameters $\boldsymbol{\theta}$, $p(\boldsymbol{\theta})$ constitutes our prior knowledge of the model parameters and $p(\boldsymbol{x})$ is the evidence of the data.

Obtaining the posterior can be computationally expensive. Traditionally, to tackle this problem we first assume a functional form for the likelihood and then apply either basic MCMC \citep[e.g.][]{Metropolis:1953p2224,Hastings:1970p97,Green:1995} or more complex nested sampling \citep[e.g.][]{Skilling:2004,Skilling:2006,Feroz:2009,Handley:2015,Buchner:2016,Buchner:2019}, where the latter has the additional benefit of providing an estimate of the evidence which is useful for performing model selection. Fundamentally, these approaches are limited by the inherent assumptions that enter into constructing the functional form of the likelihood and/or how computationally intensive the likelihood calculation becomes.

SBI on the other hand replaces the explicit evaluation of the likelihood with a stochastic simulator that provides the implicit mapping from a set of model parameters $\boldsymbol{\theta}_{i}$ to the data $\boldsymbol{x}_{i}$. These sample-parameter pairs $\left[(\boldsymbol{x}_{1},\boldsymbol{\theta}_{1}),...(\boldsymbol{x}_{N},\boldsymbol{\theta}_{N})\right]$ generated by this stochastic simulator are then used for training the neural network that either estimates the posterior directly, the likelihood or the likelihood-to-evidence ratio \citep[see e.g.][for detailed discussions on benchmarking these different approaches]{Lueckmann:2021}. Our only requirement of this simulator is that it contains a representation of the signal (realisation) combined with the noise properties, i.e. modelling or observational uncertainties. The advantage of these approaches is that since they explicitly use the output of the forward-modelled simulations rather than a likelihood evaluation we can accurately characterise the inherent variances (e.g. cosmic variance and instrumental noise) within our simulations and/or our summary statistics.

For this work, we use \textsc{\small Swyft} \citep{Miller:2022} which utilises marginal neural ratio estimation (MNRE; e.g. \citealt{Durkan:2020,Hermans:2021}) in order to learn the marginal likelihood-to-evidence ratio rather than the full likelihood-to-evidence ratio of the entire astrophysical parameter set. We denote this marginal likelihood-to-evidence ratio, $r(\boldsymbol{x},\boldsymbol{\tilde\theta})$ where $\boldsymbol{\tilde\theta}$ signifies that we are referring to an individual parameter pair (e.g $(\theta_{i}, \theta_{j}))$ rather than the full parameter set, $\boldsymbol{\theta}$. We then write this quantity, $r(\boldsymbol{x},\boldsymbol{\tilde\theta})$ as:
\begin{eqnarray}
r(\boldsymbol{x},\boldsymbol{\tilde\theta}) \equiv \frac{p(\boldsymbol{x}|\,\boldsymbol{\tilde\theta})}{p(\boldsymbol{x})} = \frac{p(\boldsymbol{\tilde\theta}|\,\boldsymbol{x})}{p(\boldsymbol{\tilde\theta})} = \frac{p(\boldsymbol{x},\boldsymbol{\tilde\theta})}{p(\boldsymbol{x})p(\boldsymbol{\tilde\theta})}.
\end{eqnarray}
This quantity is equal to the ratio of probability densities for jointly drawn sample-parameter pairs, $\boldsymbol{x},\boldsymbol{\tilde\theta} \sim p(\boldsymbol{x},\boldsymbol{\tilde\theta})$ and marginally drawn pairs $\boldsymbol{x},\boldsymbol{\tilde\theta} \sim p(\boldsymbol{x})p(\boldsymbol{\tilde\theta})$. To determine this ratio, a binary classifier is trained, $d_{\phi}(\boldsymbol{x},\boldsymbol{\tilde\theta})$, where $\phi$ denotes the network parameters. This discriminates between two hypotheses, whether the sample-parameter pairs are jointly ($C=1$) or marginally ($C=0$) drawn with both classes sampled with equal probability. This network is trained using a binary-cross entropy loss function:
\begin{eqnarray}
L[d_{\phi}(\boldsymbol{x},\boldsymbol{\tilde\theta})] &=& -\int {\rm d}\boldsymbol{x}{\rm d}\boldsymbol{\tilde\theta}\left\{ p(\boldsymbol{x},\boldsymbol{\tilde\theta}) {\rm log}d_{\phi}(\boldsymbol{x},\boldsymbol{\tilde\theta}) \right. + \nonumber \\ 
& & \left. p(\boldsymbol{x})p(\boldsymbol{\tilde\theta}){\rm log}\left[ 1- d_{\phi}(\boldsymbol{x},\boldsymbol{\tilde\theta}) \right]\right\},
\end{eqnarray}
which is minimised when $d_{\phi}(\boldsymbol{x},\boldsymbol{\tilde\theta})$ corresponds to the probability of the class labelled $C=1$ (i.e. the jointly sampled probability):
\begin{eqnarray}
d_{\phi}(\boldsymbol{x},\boldsymbol{\tilde\theta}) = p(C=1|\boldsymbol{x},\boldsymbol{\tilde\theta}) = \frac{p(\boldsymbol{x},\boldsymbol{\tilde\theta})}{p(\boldsymbol{x},\boldsymbol{\tilde\theta}) + p(\boldsymbol{x})p(\boldsymbol{\tilde\theta})} \equiv \sigma[{\rm log}\, r(\boldsymbol{x},\boldsymbol{\tilde\theta})], && \nonumber \\
&&
\end{eqnarray}
which allows the likelihood-to-evidence ratio, $r$, to be expressed in terms of the binary classifier, $d_{\phi}$ using the sigmoid function, $\sigma(y) = [1+{\rm e}^{-y}]^{-1}$.

The key advantage of this approach is that by learning the marginal likelihood-to-evidence ratio, when training this ratio we only need to provide the explicit values for the parameter combinations we are interested in. That is, for parameter inference for an $M$ dimensional model, we are only required to train $M$ 1D and $M(M-1)/2$ 2D networks to describe the full marginal posterior distribution. This is because the marginalisation over the remaining (nuisance) model parameters is implicitly included within the training data supplied by the stochastic simulator. Thus the variance due to these parameters will be included in the data but is not necessary for training the binary classifier.

\subsection{Simulated data} \label{sec:sim_data}

In order to perform SBI, we are required to have a stochastic simulator of the cosmic 21-cm signal which includes and characterises our understanding of the modelling and observational uncertainties. Below, we outline our steps to generate realistic 21-cm data.

Firstly, we use \cmfst{} to generate our 3D realisations of the cosmic 21-cm signal. These are simulated using a comoving box size of  250$^{3}$ Mpc$^{3}$ sampled on 150$^{3}$ grids using a density field which has been downsampled from a higher resolution box (450$^{3}$). We construct our 3D 21-cm light-cones spanning from $z=25$ down to $z=5.2$. In total, we generate 120,000 realisations of the cosmic 21-cm signal for each of the five different HMFs outlined earlier. Note however, to minimise the total computational overheads we only generate one set of initial conditions per set of 5 different HMFs. However, the astrophysical parameters for each of these HMFs are varied to ensure there is no overlap in the same region of parameter space with the same initial conditions.

\subsubsection{Instrumental noise}

In this work, our exploration of the impact of the HMF on our astrophysical parameters is based on a mock observation of the 21-cm PS using the SKA. To embed the instrumental noise characteristics into our mock 21-cm PS, we first split our 3D light-cone along the line-of-sight into equal comoving distance (250 Mpc) chunks. Next, since radio interferometers only observe the spatial fluctuations in the signal (zero mean signal), we measure the mean of our individual chunks and remove it from the data cubes.

To mimic the instrumental effects of the SKA on our 3D 21-cm data, we use a modified version of the publicly available \textsc{\small Python} module \textsc{\small 21cmSense}\footnote{https://github.com/jpober/21cmSense}\citep{Pober:2013p41,Pober:2014p35}. First, it generates gridded $uv$-visibilities based of any instrumental setup, for which we use the antenna configuration outlined in the SKA System Baseline Design document\footnote{http://astronomers.skatelescope.org/wp-content/uploads/2016/09/SKA-TEL-SKO-0000422\textunderscore 02\textunderscore SKA1\textunderscore LowConfigurationCoordinates-1.pdf} which corresponds to 512 37.5m antennae stations distributed within a 500m core radius. These stations are modelled assuming a system temperature, $T_{\rm sys} = 1.1T_{\rm sky} + 40~{\rm K}$ with a sky temperature of $T_{\rm sky} = 60\left(\frac{\nu}{300~{\rm MHz}}\right)^{-2.55}~{\rm K}$ \citep{Thompson2007}. We assume a total observing time of 1000 hours based on a single six-hour phase-tracked scan of the sky per night. Using these gridded $uv$-visibilities, \textsc{\small 21cmSense} then estimates the total thermal noise PS, $P_{\rm N}(k)$;
\begin{eqnarray} \label{eq:NoisePS}
P_{\rm N}(k) \approx X^{2}Y\frac{\Omega^{\prime}}{2t}T^{2}_{\rm sys},
\end{eqnarray} 
where $X^{2}Y$ is a conversion between observing bandwidth, frequency and co-moving distance, $\Omega^{\prime}$ is a beam-dependent factor derived by \citet{Parsons:2014p781} and $t$ is the total observing time. 

For the purposes of this work, we are not interested in the total thermal noise power, instead we want to directly corrupt the 3D 21-cm data cube. Therefore, following \cite{Greig:2022}, we perform the following steps:
\begin{itemize}
\item We first 3D Fourier transform the input (simulated) mean removed 21-cm data cube
\item We then filter this cube using the gridded $uv$-visibilities for the SKA computed by \textsc{\small 21cmSense}. Cells with finite $uv$-coverage are multiplied by unity, all others are set to zero
\item At each cell we then determine the amplitude of the thermal noise, $P_{\rm N}(k_{x}, k_{y}, k_{z})$, using Equation~\ref{eq:NoisePS} where $k_x$ and $k_y$ correspond to the two transverse (on sky) directions and $k_z$ is the line-of-sight direction
\item We then add random noise (zero mean with variance based on the power spectrum amplitude in the cell) to each cell to mimic the effect of thermal noise
\item Finally, we then 3D inverse Fourier transform back to obtain our noisy 21-cm data.
\end{itemize}

\subsubsection{Foreground avoidance} \label{sec:wedge}

Unfortunately, the $uv$ visibilities of radio interferometers are frequency dependent meaning that line-of-sight (frequency dependent) power can bleed into the transverse (frequency independent) Fourier modes. This `feature' manifests as a relatively well-defined contaminated `wedge' in cylindrical 2D Fourier space \citep{Datta:2010p2792,Vedantham:2012p2801,Morales:2012p2828,Parsons:2012p2833,Trott:2012p2834,Thyagarajan:2013p2851,Liu:2014p3465,Liu:2014p3466,Thyagarajan:2015p7294,Thyagarajan:2015p7298,Pober:2016p7301,Murray:2018}. Although it is theoretically possible to mitigate or `clean' these contaminated modes (see e.g.\ \citealt{Chapman:2019} for a review, or by using machine learning \citealt{Gagnon-Hartman:2021}) in this work we take the conservative approach of cutting all contaminated `wedge' modes from our 21-cm light-cones. That is, we adopt a foreground-avoidance observing strategy, whereby we only measure our 21-cm PS using `clean' Fourier modes above this `wedge'.

The boundary confining this foreground `wedge' in 2D Fourier space is given by,
\begin{eqnarray} \label{eq:wedge}
k_{\parallel} =  mk_{\perp} + b
\end{eqnarray} 
where $k_{\parallel}$ and $k_{\perp}$ are the line-of-sight and transverse Fourier modes, $b$ is a additive buffer which we assume to be $\Delta k_{\parallel} = 0.1 \,h$~Mpc$^{-1}$ which accounts for bleeding of noise extending beyond the horizon limit and $m$ is the gradient of this boundary given by
\begin{eqnarray}
m = \frac{D_{\rm C}H_{0}E(z){\rm sin}(\theta)}{c(1+z)}.
\end{eqnarray} 
This boundary depends on the comoving distance, $D_{\rm C}$, the Hubble constant, $H_{0}$, cosmological factor $E(z) = \sqrt{\Omega_{\rm m}(1+z)^{3} + \Omega_{\Lambda}}$ and ${\rm sin}(\theta)$ denotes the viewing angle of the telescope, for which we conservatively take as $\theta = \pi/2$ (i.e. a zenith pointing observation).
 
Including the imprint of foreground avoidance in our inference pipeline relies on an additional step to those outlined above. After 3D Fourier transforming our input 3D 21-cm data cube, we first zero all modes that fall below this foreground `wedge' before adding the thermal noise for all modes above the wedge.

\section{Results when incorrectly assuming a specific HMF} \label{sec:fixed}

Our first investigation is to quantify the bias in our inferred astrophysical parameters based on the assumption of an incorrectly chosen HMF model. Using our mock 21-cm observation assuming a ST HMF, we split the 3D 21-cm light-cone into ten 21-cm PS spanning from $z=5.7$ to $z=18.1$ (see e.g. Figure~\ref{fig:21cmPS}). Although the SKA is expected to observe down to 50 MHz ($z\sim27.8$), at these redshifts the thermal noise significantly dominates over the 21-cm signal. Therefore, for computational ease we restrict our data to a lower redshift ($z\leq18$). For each 21-cm PS, we only consider Fourier modes between $k=0.1$~Mpc$^{-1}$ and $k=1.0$~Mpc$^{-1}$, which leads to six Fourier modes per 21-cm PS. This lower limit loosely corresponds to when cosmic variance begins to dominate over the 21-cm signal whereas the limit at $k=1.0$~Mpc$^{-1}$ is when shot-noise in the numerical modelling becomes dominant \citep{Greig:2015p3675}.

Following \citet{Saxena:2023}, to obtain our marginal posterior distributions for our astrophysical parameters we express our 21-cm PS data as a single, 1D array (i.e. 60 data-points). This data is then the input layer to a three-layered fully connected neural network consisting of 256 neurons. The network is trained in batches of size 64, with an initial learning rate of 10$^{-3}$ that is decayed by 0.95 after each epoch. For a single HMF, the training of all the MNRE networks for our astrophysical model takes $\sim1$ hour using a single Nvidia A100. Once we have our MNRE networks, parameter inference only takes a few minutes. One of the significant advantages of MNRE is that we can more readily assess the performance of our trained network using the network convergence \citep{Cole:2022}. We provide an example of this in Appendix~\ref{sec:coverage}.

\subsection{Mock observation using only the 21-cm PS}

\begin{figure*}
	\includegraphics[trim = 0.5cm 0.35cm 0.7cm 0.5cm, scale = 0.92]{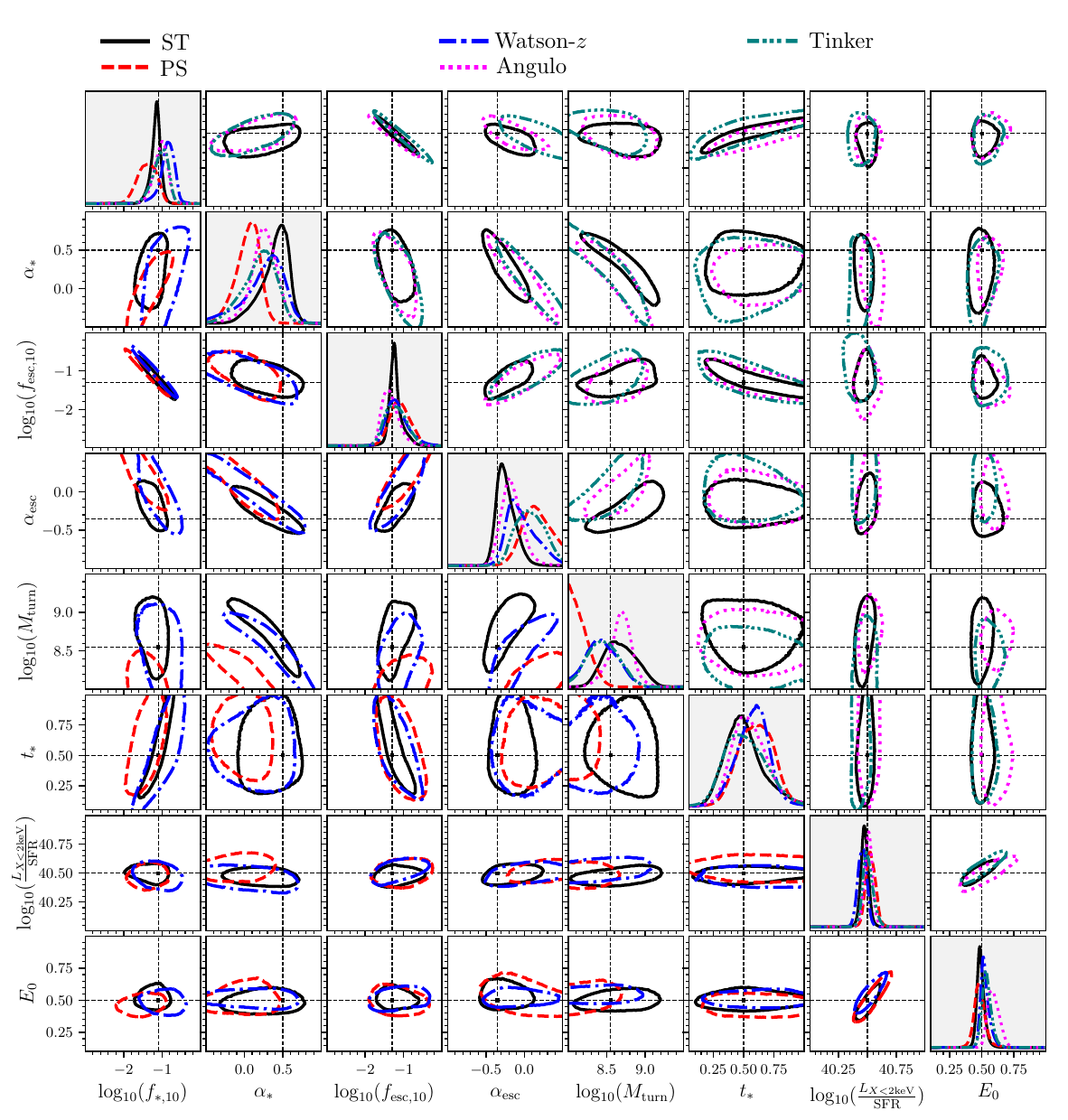}
    \caption{The recovered 1 and 2D marginalised posteriors on our astrophysical parameters assuming a mock 1000 hr observation of the 21-cm PS assuming foreground avoidance with the SKA (see Section~\ref{sec:sim_data} for further details). For this our mock observation assumes a ST HMF \citep{Sheth:2001}, while our forward modelled simulations assume a different fixed HMF. In particular, we consider PS (\citealt{Press:1974}; red-dashed), Tinker (\citealt{Tinker:2010}; teal triple dot dashed), Angulo (\citealt{Angulo:2012}; magenta-dotted) and the redshift dependant Watson (\citealt{Watson:2013}; blue-dot-dashed) HMFs. The diagonal panels correspond to the marginalised 1D posteriors whereas the off-diagonal contours correspond to the 95th percentile joint marginalised posteriors. The vertical and horizontal black dashed lines denote our fiducial astrophysical parameter set. Note, for ease of visualisation, we split the 2D posteriors above (Angulo and Tinker) and below (PS and Watson-$z$) the diagonal relative to the fiducial ST HMF.}
    \label{fig:21cmPS-noLF}
\end{figure*}

In Figure~\ref{fig:21cmPS-noLF} we present our marginalised posteriors for our 1000 hr mock observation of the 21-cm PS with the SKA, which assumes our fiducial parameter set and a ST HMF. Along the diagonal, we provide the 1D marginalised posteriors whereas above and below the diagonal we provide the 95th percentile joint marginalised posteriors. Using our mock observation, we train \textsc{\small Swyft} on our forward-modelled simulation data fixing the HMF for the same mock\footnote{Note, when initially performing our parameter inference we also trialed the SBI approach developed by \citet{Prelogovic:2023}, namely the conditional masked-autoregressive flow model to learn the actual likelihood before utilising it within an MCMC framework for inference. For completeness in Appendix~\ref{sec:CMAF} we verified the consistency of these two approaches, however, for the remainder of this work we only use \textsc{\small Swyft} for computational efficiency.}. The coloured curves correspond to assuming ST (solid, black), PS (red, dashed), Watson-$z$ (blue, dot-dashed), Angulo (magenta, dotted) and Tinker (teal, triple-dot-dashed). Finally, in Table~\ref{tab:results} we summarise the 68th marginalised percentiles for each model.

\begin{table*}
 \caption{A summary of the recovered astrophysical parameter constraints plus 68th percentiles when we vary the underlying HMF in our inference pipeline. For this, we consider a mock observation of the 21-cm PS, assuming foreground wedge avoidance and a 1000 hr observation with the SKA. Note, for our mock observation we assume a ST HMF \citep{Sheth:2001}. Additionally, we consider the recovered posteriors for our astrophysical parameters once we fold in the additional constraining power from UV LFs (see text for further details).}
 \label{tab:results}
\begin{tabular}{ccccccccc}
  \hline
 & ${\rm log}_{10}(f_{\ast,10})$ & $\alpha_{\ast}$ & ${\rm log}_{10}(f_{{\rm esc},10})$ & $\alpha_{\rm esc}$ & ${\rm log}_{10}(M_{\rm turn})$ & $t_{\ast}$ & ${\rm log}_{10}\left(\frac{L_{X<2\,{\rm keV}}}{\rm SFR}\right)$ & $E_{0}$\\
 & & & & & ($M_{\odot}$) & & (erg s$^{-1}$ $M^{-1}_{\odot}$ yr$^{-1}$) & (keV) \\
  \hline
Mock Observation & -1.1 & 0.5 & -1.30 & -0.35 & 8.55 & 0.5 & 40.5 & 0.5 \\  
  \hline
21-cm PS only &  & \\  
  \hline
ST & -1.14$\substack{+0.10\\-0.13}$  & 0.48$\substack{+0.13\\-0.19}$ & -1.25$\substack{+0.12\\-0.14}$  & -0.29$\substack{+0.14\\-0.08}$  & 8.60$\substack{+0.27\\-0.15}$  & 0.48$\substack{+0.14\\-0.15}$  & 40.47$\substack{+0.03\\-0.03}$  & 0.49$\substack{+0.03\\-0.02}$  \\
PS &  -1.37$\substack{+0.23\\-0.26}$  & 0.11$\substack{+0.11\\-0.18}$ & -1.13$\substack{+0.27\\-0.24}$  & 0.12$\substack{+0.22\\-0.15}$  & 8.00$\substack{+0.20\\-0.00}$  & 0.59$\substack{+0.16\\-0.13}$  & 40.53$\substack{+0.04\\-0.04}$  & 0.48$\substack{+0.04\\-0.04}$  \\
Tinker &  -0.98$\substack{+0.16\\-0.22}$  & 0.25$\substack{+0.19\\-0.22}$ & -1.28$\substack{+0.31\\-0.23}$  & 0.03$\substack{+0.22\\-0.16}$  & 8.44$\substack{+0.21\\-0.20}$  & 0.46$\substack{+0.20\\-0.13}$  & 40.49$\substack{+0.05\\-0.03}$  & 0.53$\substack{+0.04\\-0.03}$ \\ 
Angulo &  -0.98$\substack{+0.13\\-0.21}$  & 0.25$\substack{+0.17\\-0.14}$ & -1.42$\substack{+0.23\\-0.18}$  & -0.19$\substack{+0.15\\-0.12}$  & 8.70$\substack{+0.14\\-0.14}$  & 0.51$\substack{+0.16\\-0.12}$  & 40.51$\substack{+0.03\\-0.03}$  & 0.58$\substack{+0.05\\-0.05}$  \\
Watson-$z$ & -0.85$\substack{+0.18\\-0.15}$  & 0.37$\substack{+0.19\\-0.27}$ & -1.25$\substack{+0.27\\-0.24}$  & -0.15$\substack{+0.24\\-0.12}$  & 8.38$\substack{+0.25\\-0.16}$  & 0.61$\substack{+0.10\\-0.15}$  & 40.46$\substack{+0.03\\-0.04}$  & 0.51$\substack{+0.03\\-0.03}$ \\  
\hline
21-cm PS + UV LFs &  & \\  
  \hline
ST & -1.11$\substack{+0.13\\-0.13}$  & 0.45$\substack{+0.08\\-0.09}$ & -1.20$\substack{+0.13\\-0.11}$  & -0.30$\substack{+0.05\\-0.05}$  & 8.47$\substack{+0.16\\-0.12}$  & 0.58$\substack{+0.16\\-0.12}$  & 40.49$\substack{+0.03\\-0.02}$  & 0.49$\substack{+0.04\\-0.03}$  \\
PS &  -0.95$\substack{+0.14\\-0.21}$  & 0.52$\substack{+0.14\\-0.08}$ & -1.40$\substack{+0.15\\-0.19}$  & -0.17$\substack{+0.04\\-0.06}$  & 8.00$\substack{+0.07\\-0.00}$  & 0.59$\substack{+0.20\\-0.25}$  & 40.50$\substack{+0.04\\-0.03}$  & 0.57$\substack{+0.08\\-0.05}$  \\
Tinker &  -0.92$\substack{+0.13\\-0.17}$  & 0.48$\substack{+0.08\\-0.08}$ & -1.41$\substack{+0.19\\-0.14}$  & -0.23$\substack{+0.04\\-0.05}$  & 8.31$\substack{+0.11\\-0.15}$  & 0.43$\substack{+0.23\\-0.14}$  & 40.52$\substack{+0.03\\-0.03}$  & 0.52$\substack{+0.05\\-0.05}$  \\ 
Angulo &  -1.15$\substack{+0.11\\-0.13}$  & 0.45$\substack{+0.08\\-0.08}$ & -1.32$\substack{+0.18\\-0.09}$  & -0.20$\substack{+0.05\\-0.07}$  & 8.27$\substack{+0.16\\-0.15}$  & 0.65$\substack{+0.16\\-0.17}$  & 40.50$\substack{+0.04\\-0.03}$  & 0.52$\substack{+0.04\\-0.04}$  \\
Watson-$z$ &  -1.03$\substack{+0.12\\-0.13}$  & 0.34$\substack{+0.08\\-0.07}$ & -1.33$\substack{+0.16\\-0.16}$  & -0.15$\substack{+0.06\\-0.04}$  & 8.37$\substack{+0.18\\-0.09}$  & 0.64$\substack{+0.11\\-0.20}$  & 40.47$\substack{+0.04\\-0.03}$  & 0.49$\substack{+0.05\\-0.04}$   \\ 
  \hline
 \end{tabular}
\end{table*}

Focussing first on the ST HMF results, which serves as our baseline for the relative amplitudes of the marginalised uncertainties, we recover tight (and unbiased) constraints on our model parameters. Further, the relative uncertainties are broadly consistent with the direct MCMC approaches using \cmmc{} by \citet{Park:2019} and \citet{Greig:2020b}, albeit for a slightly different model. Therefore, as was observed by \citet{Saxena:2023}, who performed a direct comparison between the approaches, \textsc{\small Swyft} performs extremely well at obtaining robust posteriors using the 21-cm signal.

Immediately evident from Figure~\ref{fig:21cmPS-noLF} are several strong biases in the inferred astrophysical parameters owing to the incorrect HMF being assumed for our mock observation. These biases are strongest for the UV galaxies parameters, with the X-ray parameters being relatively unaffected. However, this is not too surprising given the role of the star-forming galaxies throughout the entirety of reionisation. The star-formation rate both controls the number of UV photons produced for ionising the IGM while also setting the amplitude of the production rate of X-rays which subsequently heat the IGM during the epoch of heating. Although the X-ray emissivity is proportional to the product of the X-ray luminosity and star-formation rate density (see Equation~\ref{eq:emissivity}), the dominant role of the latter during reionisation ensures that the X-ray parameters are unaffected by the choice of HMF despite the increasingly larger differences in the HMFs for increasing redshift. This same behaviour was also observed within the global 21-cm signal analysis by \citet{Mirocha:2021}. Further contributing to the reduced impact on the X-ray parameters will be the relatively larger observational uncertainties which increase for increasing redshift, where the 21-cm signal becomes more sensitive to X-ray parameters. Below, we perform an in-depth analysis of the various trends in our inferred parameters to gain further insight into the role of the HMF.

The largest biases occur for the PS model relative to our mock ST observation. For example, the inferred value of $M_{\rm turn}$ peaks at the edge of our prior region\footnote{Note, this lower bound is set by only considering atomically cooled galaxies. In order to reduce this boundary we would need to also consider a secondary population of molecularly cooled galaxies \citep[e.g.][]{Qin:2020,Qin:2021-alt}}, $M_{\rm turn} = 10^{8}\,M_{\odot}$, implying that to match the mock observation the PS HMF must produce many more low mass galaxies. This requirement is primarily set by the X-ray heating epoch, where from Figure~\ref{fig:history}, the absorption trough is delayed by $\Delta z\sim2$ relative to the ST model. Therefore, to shift to a comparable timing in order to have the 21-cm PS achieve similar amplitudes at the correct redshifts, the PS model requires many more haloes at earlier times due to the fact that the amplitude of the PS HMF drops away significantly to increasing redshift (see Figure~\ref{fig:HMF}). However, the PS HMF also rapidly produces star-forming galaxies during the EoR, as seen by the sharper gradient in the reionisation history. Therefore simply dropping $M_{\rm turn} = 10^{8}\,M_{\odot}$ would cause the EoR to occur much earlier than for the ST model. To prevent this, we are required to delay (or suppress) the escape of the UV photons from these star-forming galaxies. However, since we require star-formation to produce X-rays, the dominant way we can suppress ionisations during the EoR is through modifications to $f_{\rm esc}$ (i.e. prevent them escaping the host galaxies). For our mock observation, we have $\alpha_{\rm esc} = -0.35$ implying higher escape fractions for lower mass galaxies. Instead, if we assume a PS HMF, we recover $\alpha_{\rm esc} = 0.12$, which suppresses the escape of ionising photons by a factor of 10. Note, we also see a reduction in amplitude in ${\rm log}_{10}(f_{\ast,10})$ of $\sim20$ per cent, which corresponds to an absolute reduction in star-formation of $\sim70$ per cent. In summary, the rapidly evolving amplitude and shape of the PS HMF drives these relatively extreme biases for the astrophysical parameters.

Generally speaking for the remaining HMFs, the relative biases are less extreme, owing to their similar and more modest evolutions in the overall HMF amplitude as well as the power-law slope and exponential decay for increasing halo mass. As one would anticipate from Figure~\ref{fig:history}, both Tinker and Watson-$z$ recover similar behaviour. For these we have a delay in EoR heating of $\Delta z \sim1$, thus we recover a preference for a lower $M_{\rm turn}$ by $\sim2$ per cent (logarithmic), which albeit seemingly small corresponds to a $\sim1\sigma$ reduction or a $\sim40$ per cent lower halo mass. Additionally, we prefer a $\sim15-30$ per cent increase in ${\rm log}_{10}(f_{\ast,10})$ corresponding to an absolute increase in the star-formation of $\sim50-100$ per cent. Note, although these differences appear quite large for relatively similar models, there is a strong degeneracy between ${\rm log}_{10}(f_{\ast,10})$ and $\alpha_{\rm esc}$, thus the higher ${\rm log}_{10}(f_{\ast,10})$ for the Watson-$z$ model is compensated for by a correspondingly lower $\alpha_{\rm esc}$. These same trends also follow for the Angulo HMF, albeit in the opposite direction. For an Angulo HMF, reionisation occurs slightly earlier, thus we require a shift to a slightly higher ${\rm log}_{10}(M_{\rm turn})\sim8.70$ to slightly delay it (fewer low mass galaxies). However, this has the consequence of marginally delaying the onset of X-ray heating, which can be compensated for by increasing ${\rm log}_{10}(f_{\ast,10})$, but equally kept in balance by a slightly higher $E_{0}$ (ensuring X-ray heating is not too efficient, albeit with a larger uncertainty). To ensure the increasing ${\rm log}_{10}(f_{\ast,10})$ also does not result in early reionisation a higher ${\rm log}_{10}(f_{\rm esc,10})$ is preferred. 

In summary, the relative biases due to the choice of assumed HMF is governed by ensuring the resultant 21-cm PS matches the mock observation both during the EoR and during the X-ray heating epoch. Overall, depending on the choice of HMF, we can recover differences between the input and recovered constraints in ${\rm log}_{10}(f_{\ast,10})$ of $\pm30$ per cent, up to $\sim20$ per cent for $\alpha_{\ast}$, $\pm15$ per cent for ${\rm log}_{10}(f_{{\rm esc},10})$, up to $\sim7$ per cent for ${\rm log}_{10}(M_{\rm turn})$ and up to $\sim20$ per cent for $t_{\ast}$. In terms of the relative uncertainties from our mock ST observation, these correspond to up to 2, 2, 1, 3, 0.5$\sigma$, respectively. Note, throughout this work we define these relative biases as the separation between the maximum $a$-posteriori values for each individual parameter to the fiducial values of the ST mock observation divided by the recovered $1\sigma$ values when assuming the correct ST HMF. By far the largest impact is on $\alpha_{\rm esc}$ which controls the mass dependence of the escape fraction. Often we require more X-rays earlier, which can be achieved through an increased star-formation rate, with the consequence of causing reionisation to occur too early. Therefore the only way we can limit ionisations during the EoR is through, $f_{\rm esc}$ and principally through the mass dependence, $\alpha_{\rm esc}$. For $\alpha_{\rm esc}$ we can recover biases up to $\sim150$ per cent, which roughly corresponds to $\sim3-4\sigma$.

As noted earlier, the evolution of the 21-cm signal and summary statistics for the Angulo HMF model are extremely similar to our mock ST HMF model (see Figures~\ref{fig:history} and~\ref{fig:21cmPS}). Therefore, we can use the differences between these two models as a guide to the impact of even relatively small differences between the assumed HMF. Of our eight astrophysical parameters, four are constrained to values either right at the edge or beyond the 68th percentile boundaries. Most notably, $\alpha_{\ast}$ is $\sim2$ times lower than the input value. Therefore, even relatively modest differences between the underlying HMF can cause biases in the inferred astrophysical parameters at or beyond the inferred 68th percentile limits.

\subsection{Star-formation rate density} \label{sec:SFRD}

\begin{figure*}
	\includegraphics[trim = 0.4cm 0.3cm 0.7cm 0.5cm, scale = 0.88]{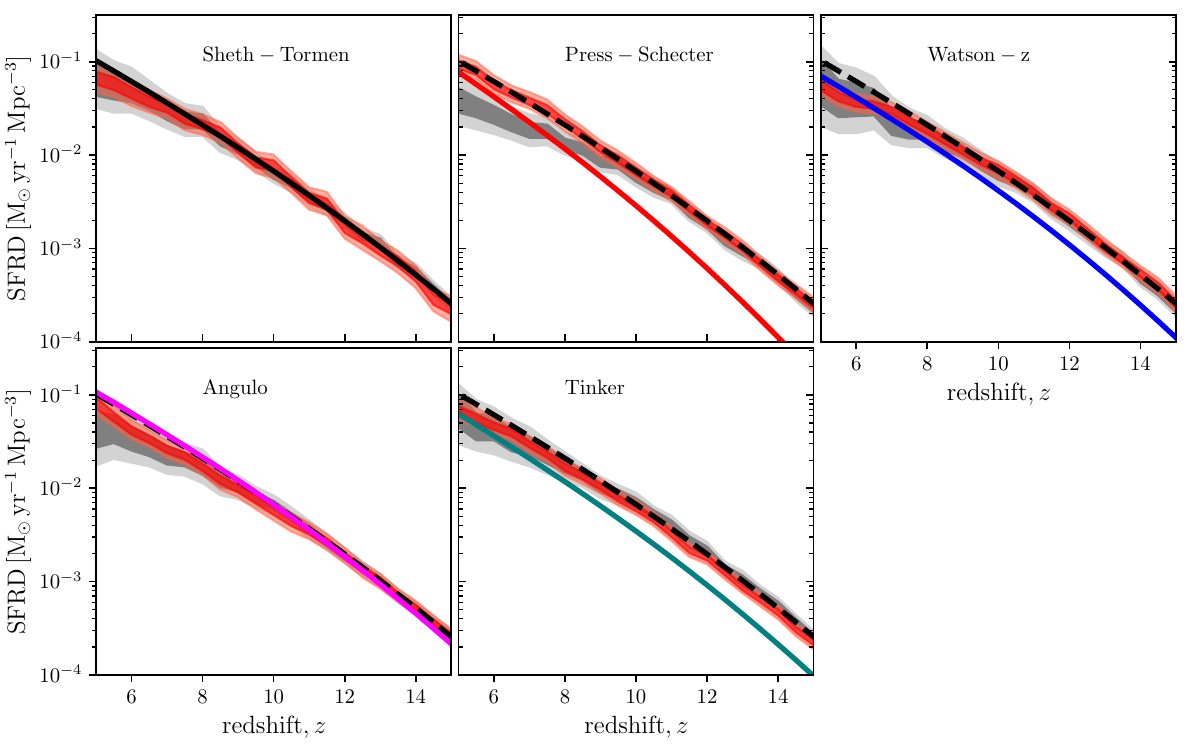}
    \caption{The marginalised posteriors on the mean star-formation rate density (SFRD) following our simulation-based inference pipeline for our mock 21-cm PS observation with the SKA. Each panel highlights the 68th (dark) and 95th (light) marginalised posteriors for an assumed HMF. The grey contours correspond to the marginalised posteriors using only the 21-cm PS and the red contours correspond to the posteriors when using both the 21-cm PS and the UV LFs. The black dashed curve in each panel is the mean SFRD for the mock observation (assuming a Sheth-Tormen HMF). The coloured curves in each panel correspond to the expected SFRD assuming the same astrophysical parameter set as our mock observation but with a different HMF.}
    \label{fig:SFRD-hmf}
\end{figure*}

In the previous section we established that the choice of HMF only impacts the UV properties of the galaxies, with the X-ray parameters left relatively unaffected (although we do observe a fairly minor bias in $E_{0}$ for the Angulo model). The relative biases in these UV galaxy parameters are driven by each model attempting to match the 21-cm PS evolution over cosmic history to that of the mock ST HMF observation. The most relevant quantity tied to the evolution of the 21-cm signal with redshift is the star-formation rate density (SFRD). Firstly, this dictates when the 21-cm signal decouples from the background CMB during the \lya{} heating epoch. It then controls the production rate of X-ray photons (see Equation~\ref{eq:emissivity}) before finally driving reionisation (in combination with the escape fraction, $f_{\rm esc}$; see Equation~\ref{eq:ioncrit2}).

To more readily demonstrate this, in Figure~\ref{fig:SFRD-hmf} we provide the recovered posteriors on the mean SFRD for each assumed HMF given our mock 21-cm PS observation with the SKA. We provide the recovered marginalised 68th (95th) percentiles as dark (light) shaded regions for each HMF along with the mean SFRD from the mock observation (black dashed curve). For reference, we also provide the mean SFRD assuming the same astrophysical parameters as our mock observation, but with each assumed HMF to visualise the necessary shift in the SFRD. Within \textsc{\small Swyft} this is simply achieved by modifying the end-points of the fully connected network architecture used for estimating the marginalised likelihood-to-evidence ratio. Retaining the same 21-cm PS data as input, we simply replace our astrophysical parameters with the SFRD values as a function of redshift. This is allowed as our training set inherently varies the underlying astrophysical parameters. This retraining of the network is only necessary as we are using MNRE, where we recover the marginalised likelihood-to-evidence ratio. If instead we obtained the full likelihood-to-evidence ratio we could simply sample the full likelihood-to-evidence ratio within an MCMC framework to obtain the posteriors on the SFRD.

As expected, by assuming an incorrect HMF, the inferred astrophysical parameters that we recovered in the previous section are biased to ensure we reproduce the correct SFRD from our mock 21-cm observation. This is readily demonstrated by the shift away from the expected SFRD for each HMF (coloured curves) to the tight posteriors around the SFRD of the mock (black dashed curves). Adopting a PS, Watson-$z$ or Tinker HMF under predicts the mean SFRD, therefore the inferred astrophysical parameters we recovered previously were biased to increase the overall SFRD. For example, we inferred lower $M_{\rm turn}$ for these models and in general an increased $f_{\ast}$ which is consistent with this trend. For the Angulo HMF, it marginally over predicts and thus we inferred biased astrophysical parameters which reduce the overall stellar output, for example our observed increase in $M_{\rm turn}$ (i.e. less star-forming galaxies).

For decreasing redshift, $z<8$, the inferred mean SFRD tends to drop below the expectation from the mock observation (black solid curve). However, this is simply due to the strong degeneracy between star-formation, $f_{\ast}$, and the escape fraction, $f_{\rm esc}$ (see e.g. Figure~\ref{fig:21cmPS-noLF}). Here, when considering the SFRD we are only considering $f_{\ast}$, whereas reionisation is driven by the product of $f_{\ast}$ and $f_{\rm esc}$ (Equation~\ref{eq:ioncrit2}). For example, even when we assumed the correct ST HMF, we observe a slight decrease in the inferred mean SFRD relative to the mock observation. From Table~\ref{tab:results}, we recovered a slightly reduced ${\rm log}_{10}(f_{\ast,10})$ ($-1.14$ compared to the input $-1.1$) and correspondingly higher ${\rm log}_{10}(f_{\rm esc,10})$ ($-1.25$ compared to the input $-1.30$) along with a less negative slope for $\alpha_{\rm esc}$ ($-0.29$ compared to the input $-0.35$), which results in the slight decrease in the inferred SFRD during the EoR. This same trend additionally occurs for the other HMFs albeit with a larger amplitude owing to the different properties of the underlying HMF. For the PS HMF, the discrepancy is much larger, which corresponds to the preference for an $M_{\rm turn}$ right at the edge of our prior range which translates to much larger biases in the other galaxy parameters.

In addition to this relative downturn in the amplitude of the inferred mean SFRD at $z<8$ we also observe an increase in the width of the corresponding posteriors. This increasing width is also due to the degeneracy between $f_{\ast}$ and $f_{\rm esc}$. At high redshifts, prior to reionisation, the relative amplitude of the 21-cm signal (and statistics such as the PS) are set by X-ray heating, which are not sensitive to $f_{\rm esc}$ (their effective escape fraction is instead controlled by $E_{0}$). As a result, the posteriors at higher redshift, where the 21-cm PS is at its largest amplitude are tighter. 

\subsection{Mock 21-cm PS plus UV LF observation} \label{sec:addUVLFs}

\begin{figure*}
	\includegraphics[trim = 0.5cm 0.3cm 0.7cm 0.4cm, scale = 0.92]{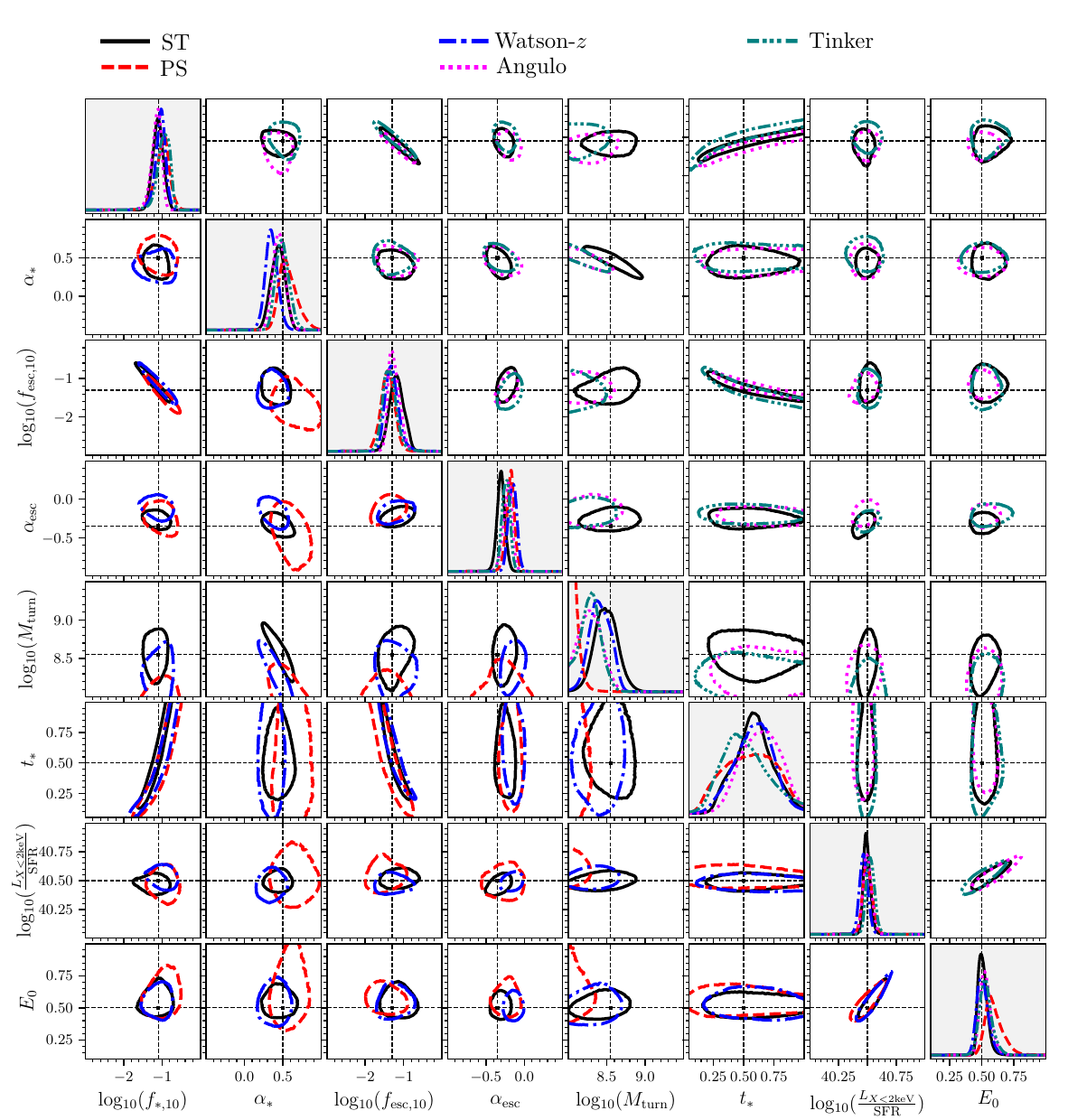}
    \caption{The same as Figure~\ref{fig:21cmPS-noLF} except also including the UV LFs at $z=6, 7, 8$ and~10 as observational priors.}
    \label{fig:21cmPS-wLF}
\end{figure*}

Thus far we have only explored the relative bias in our inferred parameters for a mock 21-cm PS observation. However, when inferring observations of the 21-cm signal, independent constraining power is available through complimentary observations. Most notably, through the addition of UV LFs \citep{Park:2019}. As outlined in Section~\ref{sec:UVLFs}, UV LFs serve to break the degeneracy between $f_{\ast}$ and $f_{\rm esc}$ by providing independent constraints on $f_{\ast}$. Therefore, in this section, we explore the impact of our choice of HMF on our inferred parameters when combining our mock 21-cm PS measurement with the SKA and existing UV LF measurement. Namely, we consider UV LFs at $z=6, 7, 8$ and 10.

In Figure~\ref{fig:21cmPS-wLF} we provide the marginalised posteriors for our 1000 hr mock observation of the 21-cm PS from the SKA combined with our 4 UV LFs. Along the diagonal, we provide the 1D marginalised posteriors whereas above and below the diagonal we provide the 95th percentile joint marginalised posteriors. The coloured curves denote the inferred astrophysical parameters for different assumed HMFs fit to the same mock observation (ST HMF). Again, we provide the marginalised 68th percentiles in Table~\ref{tab:results}.

Importantly, for SBI we require a simulator of the forward-modelled data which includes the dominant sources of uncertainty. In \cmfst{} the output UV LFs are computed analytically. Therefore, to include noise in our simulated UV LFs we simply add random noise, $\mathcal{N}(0,\sigma)$ into each corresponding $M_{\rm UV}$ bin according to the relative observational uncertainty, $\sigma(M_{\rm UV})$. Fundamentally, this assumes each $M_{\rm UV}$ bin is an independent measurement, which should be a reasonable assumption. In Appendix~\ref{sec:LF} we verify this approach reproduces the expected marginalised posteriors in comparison to a simple MCMC when considering only UV LFs. Further, our mock astrophysical model is based on the results of \citet{Qin:2021}, which deviates from the best-fit model of \citet{Park:2019} which is constrained against the same UV LFs used in this work. Therefore, in order to avoid biasing our results due to the different astrophysical parameterisations, we simply rescale the observed UV LFs by the difference between our fiducial parameter set and that of \citet{Park:2019} to ensure our mock will be consistent with the input UV LFs. Since we are focussed on quantifying the biases owing to the choice of HMF, and not constraining against real-world data, this approach is justified.

Firstly, considering our fiducial model (ST HMF), one can clearly see the impact that the UV LFs have on improving the constraining power on the UV galaxy properties. Primarily, we see significant improvements in $f_{\ast,10}$, $\alpha_{\ast}$, as these are the two most strongly constrained galaxy parameters from the UV LFs (see e.g. Figure~\ref{fig:LFcomp})). Further, as this is independent information, the inclusion of UV LFs also serve to break the degeneracies between $f_{\ast}$ and $f_{\rm esc}$ as can be clearly seen both by the notably reduced joint posterior volumes and the orientation (change in slope) of the posteriors contours. Quantitatively, following the inclusion of the UV LFs we observe a reduction in the uncertainties of $\sim25$ per cent for ${\rm log}_{10}(f_{\rm esc,10})$ with the most significant improvements in both $\alpha_{\rm esc}$ and $\alpha_{\ast}$ of $\sim50$ per cent. These improvements are consistent with those seen by \citet{Park:2019}.

Relative to Figure~\ref{fig:21cmPS-noLF}, the maximum $a$-posteriori values of the astrophysical parameters for the different incorrectly chosen HMFs has been reduced relative to the true fiducial parameters. Primarily, this reduction stems from the shrinking posterior volumes that come with the breaking of the astrophysical parameter degeneracies following the inclusion of the UV LFs. This now significantly limits the available posterior space to achieve the correct underlying SFRD to match that of the mock observation. To demonstrate this, we again use \textsc{Swyft} to obtain our marginalised posteriors on the SFRD when combining UV LFs with our mock 21-cm observation. In Figure~\ref{fig:SFRD-hmf}, the dark (light) red shaded regions denote the 68th (95th) percentiles for our joint UV LFs and 21-cm mock observation. Quite clearly, one can see significant reductions in the relative widths of the posteriors along with less of a significant downturn in the marginalised posteriors during reionisation (e.g. at $z\lesssim8$). The removal of this downturn, along with the notably reduced posterior widths corresponds to the breaking of the degeneracy between $f_{\ast}$ and $f_{\rm esc}$. Now, with $f_{\ast}$ strongly constrained by the UV LFs, there is now less posterior volume with which $f_{\rm esc}$ can achieve reionisation comparably to the mock observation. 

For the PS HMF, we again prefer $M_{\rm turn}$ right at the edge of our prior range ($10^{8}$), however, this is now inconsistent with our assumed value of 8.55 by $\sim8\sigma$. Previously to compensate for this notably reduced $M_{\rm turn}$ we required a reduced $f_{\ast,10}$ and an $f_{\rm esc}$ parameterisation which significantly reduced the escape fraction in low mass galaxies. By including UV LFs, we can no longer significantly reduce the stellar component as we would incorrectly match the observed UV LFs, thus we can only modify $f_{\rm esc}$. Because of these limitations, we now prefer a considerably different $f_{\rm esc}$, with lower ${\rm log}_{10}(f_{{\rm esc},10})$, -1.4 compared to -1.13 previously, and a decreasing halo mass dependence. As a consequence of this enforced higher star-formation rate we would overproduce the number of X-rays, however, we now observe a bias in $E_{0}$ shifting to a higher value ($\sim1.5\sigma$) to reduce the number of X-rays capable of escaping the host galaxy.

Like previously, for the remaining HMFs the behaviour of the biases are less extreme. For the Tinker and Watson-$z$ HMFs we were required to increase the relative number of UV photons to ensure reionisation could occur sooner, resulting in an overall increase of $f_{\ast}$ (primarily for lower-mass haloes) and a decrease in $M_{\rm turn}$. Following the inclusion of the UV LFs, we are now restricted by the number of UV photons we can produce, which is compensated for by a slight further reduction in $M_{\rm turn}$ to increase the number of sources (especially for the Tinker HMF). Since we are no longer producing as many stars per stellar baryon for the low-mass galaxy end, we must equivalently increase $f_{\rm esc}$ for these galaxies to ensure enough of these UV photons are entering the IGM to allow reionisation to occur earlier. For the Angulo HMF model, previously we required increasing $M_{\rm turn}$ and lowering $f_{\rm esc}$ to slightly delay the onset of reionisation and equally a higher $f_{\ast}$ to produce a sufficient number of X-rays. Including UV LFs restricts our ability to increase $f_{\ast}$, thus instead we are required to reduce $M_{\rm turn}$. This change to a lower $M_{\rm turn}$ results in a notable change in the $f_{\ast}$ mass dependence (requiring fewer stars for lower mass galaxies) whose change is balanced by an increase in $f_{\rm esc}$ to ensure the timing of reionisation is correct.

In summary, the relative amplitudes of the biases in the astrophysical parameters has drastically reduced as we no longer have the freedom to vary $f_{\ast}$ to any value/mass dependence as it is now strongly constrained by the UV LFs. As a consequence, this results in a significant reduction in the available posterior volumes for our model parameters which limits the corresponding biases. Nevertheless, although the relative amplitudes of the biases away from the fiducial parameters have reduced, quantitatively strong biases remain owing to the significantly reduced marginalised uncertainties provided by the UV LFs. For example, we now recover differences in ${\rm log}_{10}(f_{\ast,10})$ of $\sim20$ per cent, $\sim30$ for $\alpha_{\ast}$, $\sim10$ per cent for ${\rm log}_{10}(f_{{\rm esc},10})$, up to $\sim7$ per cent for ${\rm log}_{10}(M_{\rm turn})$ and $\sim30$ per cent for $t_{\ast}$. Relative to the marginalised uncertainties these correspond to up to 1.5, 2, 1, 7, 1$\sigma$, respectively. For $\alpha_{\rm esc}$ we recover much smaller biases of $\Delta \alpha_{\rm esc} \sim 0.20$ compared to $\Delta \alpha_{\rm esc} \sim 0.50$ previously, however, this still corresponds to $\sim3-4\sigma$ bias owing to the factor $\sim2-3$ reduction in the corresponding marginalised uncertainties.

Again, we can use the Angulo model as an illustrative example of the relative biases for even modest differences in the HMF away from the mock observation. Following the inclusion of the UV LFs, now only two parameters are in excess of the 68th marginalised uncertainties relative to the fiducial model compared to the four previously when only considering the 21-cm PS. However, the two inconsistent parameters, $\alpha_{\rm esc}$ and $M_{\rm turn}$ are more significantly biased (owing to the restriction of $f_{\ast}$ provided by the UV LF data). For example, $\alpha_{\rm esc}$ is recovered to $\sim -0.2$ which is now $\sim3\sigma$ from the input fiducial value (-0.35) while $M_{\rm turn}$ is $\sim2\sigma$ below the fiducial value. Therefore, even for relatively modest differences in the HMF, the inferred astrophysical parameters can be strongly biased.

\section{Jointly recovering the HMF and astrophysical parameters} \label{sec:general}

Thus far, we have explored the relative biases in the inferred model parameters owing to incorrectly assuming the HMF for our underlying astrophysical model. Generally speaking, this assumption of a fixed HMF stems from the reduced uncertainty in the HMF models relative to those describing the astrophysical properties of the galaxies allowing us to limit the computational complexity for our inference pipelines. In this section, we explore the consequences of relaxing this assumption and attempt to jointly constrain both the astrophysical parameters as well as the underlying model for the HMF\footnote{Alternatively, rather than expanding the parameter set to also include the HMF parameters, if one was only interested in the astrophysical parameters one could instead consider performing Bayesian model averaging of the recovered posteriors obtained from assuming a broad range of HMFs from the literature.}. For this, we assume the same mock 21-cm PS observation assuming a ST HMF.

\subsection{Generalised HMF} 

Throughout the literature there are a large number of different analytic fitting functions that have been derived to fit the HMFs of $N$-body simulations \citep[e.g.][]{Jenkins:2001,Warren:2006,Reed:2007,Tinker:2008,Tinker:2010,Angulo:2012,Watson:2013,Diemer:2020}. These can sometimes contain several parameters which may or may not have complex scalings with redshift and/or cosmological parameters. Deriving a universal, generalised HMF to describe all $N$-body results is beyond the scope of this work, however, we can draw from the more common parameterisations in the literature to provide a simple, basic example. 

The HMF can simply be defined as,
\begin{eqnarray} \label{eq:GeneralHMF}
\frac{{\rm d}n}{{\rm d}M} = f(\sigma)\frac{\bar{\rho}_m}{M}\frac{d{\rm ln}\sigma^{-1}}{dM}.
\end{eqnarray}
where $\sigma$ is the root mean square linear overdensity of the density field smoothed by a top-hat filter of radius, $R$, containing a mass $M$ at the mean cosmic matter density, $\bar{\rho}_m$. Defined in this way, the function $f({\sigma})$ encapsulates the intrinsic behaviour for any assumed model of the HMF.

Since our mock observation is generated assuming a ST HMF, we first consider a generalised five parameter model of the following form:
\begin{eqnarray} \label{eq:5param}
f(\sigma) = a \left[ \left(b\nu\right)^c + 1 \right]\nu^e{\rm exp}\left(-d\nu^{2}\right),
\end{eqnarray}
where $\nu = \delta_{\rm c}/\sigma$ and $\delta_{\rm c}$ is the critical density. Defined in this way, we have five free parameters which we simply denote $a$, $b$, $c$, $d$ and $e$, which is more flexible than the actual ST HMF. As a result, this form slightly differs from the actual ST HMF, whereby their HMF is parameterised assuming a $\nu^{\prime} = \alpha\delta_{\rm c}/\sigma$ in place of our $\nu$, which results in slightly different amplitudes for the various coefficients. Nevertheless, adopting $a = 0.2406$, $b = 0.85$, $c = -0.35$, $d = 0.365$ and $e = 1.0$ can mimic the behaviour for the ST HMF model. Generally speaking, this functional form can recover most HMFs in the literature, with the largest discrepancies being those models with coefficients with redshift dependence \citep[e.g.][]{Reed:2007,Tinker:2008,Tinker:2010,Watson:2013}.

While this five parameter HMF should be able to recover the ST HMF, it will also be illustrative to explore the relative impact of the assumed functional form of the HMF on our inferred parameters. That is, if we assume an incorrect function form incapable of mimicking the true form of the HMF, how significant will the relative biases be? Therefore, we shall consider a secondary model whereby we set $e=0$ in Equation~\ref{eq:5param}, thus considering a simpler four parameter generalised HMF. This choice is somewhat arbitrary as we could theoretically remove any model parameter, however, the removal of the $\nu$ multiplicative term should have a more notable impact on the HMF shape and amplitude and correspondingly the astrophysical parameters. In future, one could envisage using Bayesian evidence to more robustly determine better generalised functional forms for characterising the HMF.

\subsection{Mock observation using only the 21-cm PS}

\begin{figure*}
	\includegraphics[trim = 0.8cm 0.2cm 0.7cm 0.4cm, scale = 0.89]{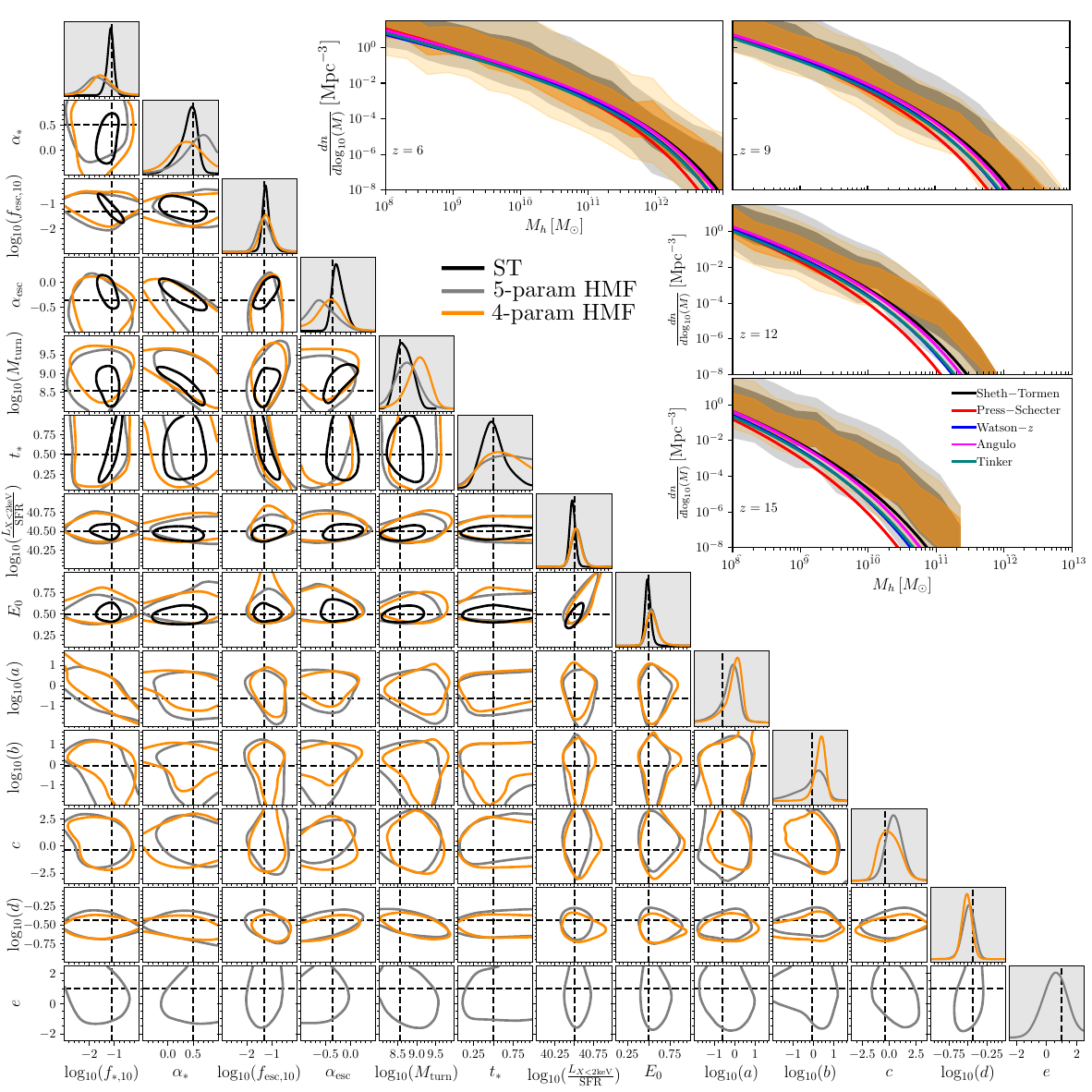}
    \caption{The recovered one and two dimensional marginalised posteriors on our joint astrophysical and general HMF function parameters assuming a mock 1000 hr observation of the 21-cm PS assuming foreground avoidance with the SKA. Note, our mock observation assumes the Sheth-Tormen HMF. The grey (orange) contours correspond to a five (four) parameter HMF (see text for further details). The vertical and horizontal black dashed lines denote our fiducial astrophysical parameter set and the corresponding values for the Sheth-Tormen HMF.
 Above the diagonal, we provide the marginalised posteriors on the halo mass function when simultaneously constraining both the astrophysical and halo mass function parameters assuming our same mock observation. Each panel corresponds to $z=6, 9, 12$ and~15, respectively. Within each panel we show the 68th (dark) and 95th (light) marginalised posteriors on the recovered HMF. The grey (orange) contours correspond to the marginalised posteriors assuming our five (four) parameter halo mass function. For reference, we overlay several different HMFs as explored earlier. The vertical offset between the marginalised posteriors and the analytic expressions is due to the degeneracy between the HMF normalisation, $a$, and the normalisations of $f_{\ast}$ and $f_{\rm esc}$ (see text for further details).
    }
    \label{fig:GeneralHMF-noLF}
\end{figure*}

\begin{table*}
 \caption{A summary of the recovered parameter constraints plus 68th percentiles when we additionally vary the underlying HMF along with the astrophysical parameters within our inference pipeline. For this, we consider the same mock observation of the 21-cm PS, assuming foreground wedge avoidance and a 1000 hr observation with the SKA and a ST HMF \citep{Sheth:2001}. Additionally, we provide the recovered constraints when also including the additional constraining power from UV LFs (see text for further details).}
 \label{tab:results-general}
\begin{tabular}{ccccccccccccc}
 \hline
Astrophysical & ${\rm log}_{10}(f_{\ast,10})$ & $\alpha_{\ast}$ & ${\rm log}_{10}(f_{{\rm esc},10})$ & $\alpha_{\rm esc}$ & ${\rm log}_{10}(M_{\rm turn})$ & $t_{\ast}$ & ${\rm log}_{10}\left(\frac{L_{X<2\,{\rm keV}}}{\rm SFR}\right)$ & $E_{0}$\\
parameters & & & & & ($M_{\odot}$) & & (erg s$^{-1}$ $M^{-1}_{\odot}$ yr$^{-1}$) & (keV) \\
 \hline
 Mock Observation  & -1.1 & 0.5 & -1.30 & -0.35 & 8.55 & 0.5 & 40.5 & 0.5 \\  
\hline
21-cm PS only &  & \\  
\hline
ST HMF & -1.14$\substack{+0.10\\-0.13}$  & 0.48$\substack{+0.13\\-0.19}$  & -1.25$\substack{+0.12\\-0.14}$ & -0.29$\substack{+0.14\\-0.08}$ & 8.60$\substack{+0.27\\-0.15}$ & 0.48$\substack{+0.14\\-0.15}$ & 40.47$\substack{+0.03\\-0.03}$ & 0.49$\substack{+0.03\\-0.02}$ \\
5-param HMF & -1.83$\substack{+0.48\\-0.47}$  & 0.66$\substack{+0.29\\-0.24}$ & -1.25$\substack{+0.30\\-0.23}$ & -0.57$\substack{+0.30\\-0.21}$ & 8.82$\substack{+0.31\\-0.27}$ & 0.64$\substack{+0.32\\-0.16}$ & 40.54$\substack{+0.06\\-0.05}$ & 0.53$\substack{+0.08\\-0.05}$ \\
4-param HMF & -1.47$\substack{+0.48\\-0.43}$  & 0.30$\substack{+0.36\\-0.29}$ & -1.23$\substack{+0.23\\-0.24}$ & -0.37$\substack{+0.28\\-0.20}$ & 9.08$\substack{+0.28\\-0.28}$ & 0.56$\substack{+0.28\\-0.18}$ & 40.50$\substack{+0.06\\-0.05}$ & 0.54$\substack{+0.07\\-0.07}$ \\
5-param (joint norm) & ---  & 0.72$\substack{+0.25\\-0.23}$ & -1.23$\substack{+0.28\\-0.27}$ & -0.65$\substack{+0.26\\-0.22}$ & 8.68$\substack{+0.30\\-0.29}$ & 0.71$\substack{+0.25\\-0.30}$ & 40.54$\substack{+0.06\\-0.06}$ & 0.54$\substack{+0.05\\-0.05}$ \\
4-param (joint norm) & ---  & 0.27$\substack{+0.36\\-0.22}$ & -1.26$\substack{+0.25\\-0.24}$ & -0.39$\substack{+0.19\\-0.27}$ & 9.24$\substack{+0.21\\-0.30}$ & 0.41$\substack{+0.40\\-0.12}$ & 40.55$\substack{+0.06\\-0.06}$ & 0.52$\substack{+0.05\\-0.04}$ \\
\hline
21-cm PS + UV LFs & \\  
\hline
ST HMF & -1.11$\substack{+0.13\\-0.13}$  & 0.45$\substack{+0.08\\-0.09}$ & -1.20$\substack{+0.13\\-0.11}$  & -0.30$\substack{+0.05\\-0.05}$  & 8.47$\substack{+0.16\\-0.12}$  & 0.58$\substack{+0.16\\-0.12}$  & 40.49$\substack{+0.03\\-0.02}$  & 0.49$\substack{+0.04\\-0.03}$ \\
5-param HMF & -1.68$\substack{+0.41\\-0.50}$  & 0.57$\substack{+0.11\\-0.09}$ & -1.06$\substack{+0.22\\-0.17}$ & -0.48$\substack{+0.12\\-0.09}$ & 8.60$\substack{+0.36\\-0.27}$ & 0.59$\substack{+0.25\\-0.22}$ & 40.53$\substack{+0.06\\-0.05}$ & 0.55$\substack{+0.07\\-0.06}$ \\
4-param HMF & -1.42$\substack{+0.56\\-0.50}$  & 0.68$\substack{+0.12\\-0.13}$ & -1.04$\substack{+0.22\\-0.18}$ & -0.55$\substack{+0.15\\-0.19}$ & 8.88$\substack{+0.24\\-0.25}$ & 0.50$\substack{+0.20\\-0.22}$ & 40.56$\substack{+0.06\\-0.08}$ & 0.53$\substack{+0.06\\-0.04}$ \\
5-param (joint norm) & ---  & 0.55$\substack{+0.18\\-0.13}$ & -1.27$\substack{+0.24\\-0.24}$ & -0.40$\substack{+0.14\\-0.12}$ & 8.77$\substack{+0.33\\-0.24}$ & $0.54\substack{+0.22\\-0.20}$ & 40.50$\substack{+0.07\\-0.06}$ & 0.54$\substack{+0.08\\-0.06}$ \\
4-param (joint norm) & ---  & 0.33$\substack{+0.14\\-0.09}$ & -1.01$\substack{+0.19\\-0.24}$ & -0.44$\substack{+0.15\\-0.21}$ & 9.08$\substack{+0.23\\-0.27}$ & 0.46$\substack{+0.28\\-0.18}$ & 40.54$\substack{+0.06\\-0.06}$ & 0.52$\substack{+0.06\\-0.05}$ \\
\hline
\\
\hline
HMF parameters & & ${\rm log}_{10}(a\times f_{\ast,10})$ & ${\rm log}_{10}(a)$ & ${\rm log}_{10}(b)$ & c & ${\rm log}_{10}(d)$ & e \\  
\hline
Mock Observation  & & -1.72 & -0.62 & -0.07 & -0.35 & -0.44 & 1.0  \\  
Prior ranges  & & [-5.00,1.70] & [-2.00,1.70] & [-2.00,1.70] & [-3.50, 3.50] & [-1.00,0.00] & [-2.50,2.50]  \\  
\hline
21-cm PS only &  \\  
\hline
5-param HMF &  & --- & -0.16$\substack{+0.36\\-0.33}$ & 0.24$\substack{+0.44\\-0.37}$ & 0.50$\substack{+0.84\\-0.84}$ & -0.47$\substack{+0.07\\-0.07}$ & 0.57$\substack{+0.85\\-0.87}$ &  \\
4-param HMF &  & --- & 0.15$\substack{+0.23\\-0.39}$ & 0.42$\substack{+0.22\\-0.31}$ & -0.43$\substack{+1.03\\-0.63}$ & -0.52$\substack{+0.06\\-0.05}$ & --- &  \\
5-param (joint norm) &  & -1.69$\substack{+0.36\\-0.39}$ & --- & 0.18$\substack{+0.49\\-0.65}$ & 0.76$\substack{+0.70\\-0.93}$ & -0.47$\substack{+0.05\\-0.08}$ & 0.21$\substack{+0.70\\-0.88}$ &  \\
4-param (joint norm) &  & -1.45$\substack{+0.27\\-0.29}$ & --- & 0.33$\substack{+0.20\\-0.26}$ & 0.46$\substack{+0.56\\-1.17}$ & -0.54$\substack{+0.05\\-0.06}$ & --- &  \\
\hline
21-cm PS + UV LFs & \\  
\hline
5-param HMF & & --- & -0.27$\substack{+0.23\\-0.35}$ & 0.02$\substack{+0.57\\-0.58}$ & 0.13$\substack{+0.79\\-0.86}$ & -0.51$\substack{+0.06\\-0.05}$ & 0.27$\substack{+0.82\\-0.97}$ & \\
4-param HMF & & --- & -0.29$\substack{+0.28\\-0.30}$ & 0.31$\substack{+0.26\\-0.70}$ & -0.43$\substack{+1.46\\-0.68}$ & -0.59$\substack{+0.07\\-0.06}$ & --- &  \\
5-param (joint norm) &  & -1.82$\substack{+0.35\\-0.31}$ & --- & 0.13$\substack{+0.49\\-0.63}$ & 0.15$\substack{+0.77\\-0.68}$ & -0.49$\substack{+0.06\\-0.05}$ & 0.86$\substack{+0.58\\-1.15}$ &  \\
4-param (joint norm) &  & -1.69$\substack{+0.31\\-0.24}$ & --- & 0.35$\substack{+0.23\\-1.21}$ & -0.62$\substack{+1.07\\-0.68}$ & -0.58$\substack{+0.05\\-0.05}$ & --- &  \\
\hline
 \end{tabular}
\end{table*}

Since we have expanded the dimensionality of our model to 13 individual parameters, we are required to generate a new database of models. Specifically, we generated a new database of 200,000 models for both the four and five parameter generalised form of the HMF. For the five new HMF parameters we adopt relatively broad priors which are summarised in Table~\ref{tab:results-general}. In Figure~\ref{fig:GeneralHMF-noLF} we demonstrate our marginalised posteriors for our model parameters using \textsc{Swyft}. Along the diagonal we show the marginalised 1D posteriors whereas below the diagonal we provide the 95th percentile joint marginalised posteriors. In each, the grey (orange) curves correspond to the recovered posteriors assuming the five (four) parameter generalised form for the HMF. For reference, we equally show the marginalised posteriors when assuming the correct ST HMF (black curves, Figure~\ref{fig:21cmPS-noLF}). Finally, in Table~\ref{tab:results-general} we summarise the marginalised constraints and 68th percentile uncertainties for each of the 13 model parameters.

As one would expect, by simultaneously constraining the astrophysical and HMF parameters, we now recover considerably broader posteriors along with several additional strong model parameter degeneracies. The most significant of these degeneracies is between the various normalisation quantities, e.g. ${\rm log}_{10}(f_{\ast,10})$, ${\rm log}_{10}(f_{\rm esc,10})$ and now the HMF normalisation, $a$. As these are multiplicative within our 21-cm model, the enlarged posterior volumes owing to these degeneracies play a role in broadening the overall posteriors of all the model parameters. Realistically, rather than attempting to constrain all three normalisation quantities independently, one could parameterise the product of ${\rm log}_{10}(a \times f_{\ast,10})$ to minimise the relative strength of these degeneracies, keeping $f_{\rm esc}$ independent as the $f_{\ast}-f_{\rm esc}$ degeneracy can be reduced following the inclusion of UV LFs. Nevertheless, as a first step we keep these three parameters independent to explore the relative amplitude of these uncertainties when attempting to constrain all model parameters.

Focussing first on the five parameter HMF model, we can correctly recover almost all of our astrophysical parameters to within the much broader 68th percentile uncertainties, with $f_{\ast,10}$ being consistent within the 95th percentile uncertainties ($M_{\rm turn}$ is right at the edge of the 68th percentile). The significant deviation in $f_{\ast,10}$ comes from the noted degeneracy between the HMF normalisation, $a$ and ${\rm log}_{10}(f_{\ast,10})$ and ${\rm log}_{10}(f_{\rm esc,10})$. For example, our recovered constraint on ${\rm log}_{10}(a)$ is -0.16 relative to the fiducial value of -0.62 ($\sim1.5\sigma$ larger). To compensate for the larger $a$, we must significantly reduce ${\rm log}_{10}(f_{\ast,10})$ (-1.83 instead of our assumed -1.1, or approximately $\sim1.5\sigma$ lower). The larger than expected ${\rm log}_{10}(f_{{\rm esc},10})$ corresponds to the fact that we have a smaller stellar component, requiring a higher escape fraction to ensure reionisation occurs at the correct point in time. For the HMF parameters, other than the strongly offset $a$, the remaining parameters are within their 68th percentiles, owing to the fact that the relative uncertainties are quite large. Interestingly, the parameter $d$ is quite strongly constrained compared to all other HMF parameters, which corresponds to the location and amplitude of the exponential drop-off within the HMF. Amongst the various HMF parameters, there are no significantly strong degeneracies other than that between $b$ and $c$.

For the four parameter HMF, we recover relatively comparable constraints on our astrophysical parameters as the five parameter model. The largest differences occur for $M_{\rm turn}$, $\alpha_{\ast}$ and $\alpha_{\rm esc}$. For $\alpha_{\ast}$ and $\alpha_{\rm esc}$ we are simply on the opposite sides of the strong $\alpha_{\ast}-\alpha_{\rm esc}$ degeneracy relative to the five parameter HMF model. However, for the four parameter model we prefer a much higher $M_{\rm turn}$ ($\sim9.08$) which is due to the strong degeneracy between $M_{\rm turn}$ and the exponential term in the HMF, $d$. As this term, $d$, is relatively strongly constrained and degenerate with $M_{\rm turn}$, its offset from the expected value leads to the observed increase in $M_{\rm turn}$.

Overall, for the astrophysical parameters the relative amplitude of the uncertainties increase due to the simultaneous constraining of the HMF parameters by a factor of $\sim4$ for ${\rm log}_{10}(f_{\ast,10})$ and $\sim2-3$ for the remainder. Relatively speaking for all parameters in our expanded model, the four parameter HMF model has slightly reduced marginalised uncertainties owing to the fact that we have one less parameter in our model. As a consequence, this will marginally amplify any respective biases if the assumed form of the HMF is unable to match that of the input ST HMF. Indeed, for the four parameter HMF model we recover a larger than expected value for $M_{\rm turn}$ (over $\sim2\sigma$) to compensate for the bias in $d$. 

\begin{figure*}
	\includegraphics[trim = 0.5cm 0cm 0cm 0.5cm, scale = 0.43]{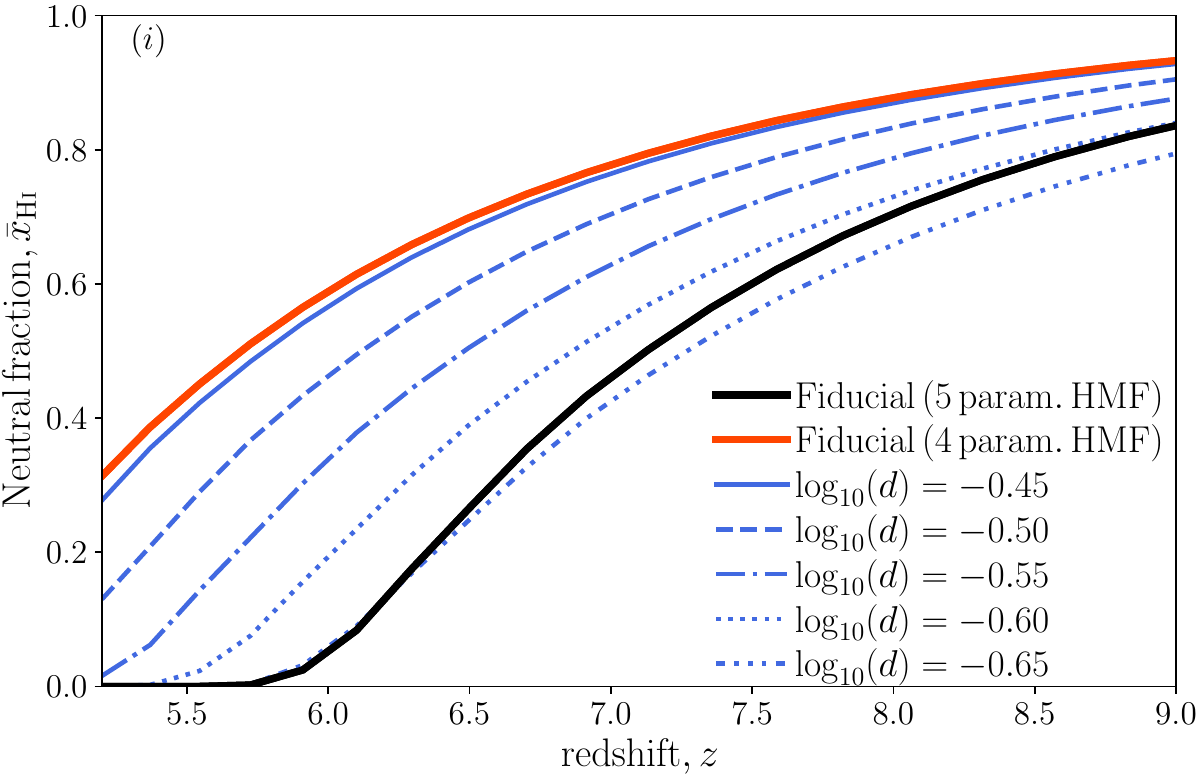}
	\includegraphics[trim = 0.2cm 0cm 0cm 0.5cm, scale = 0.43]{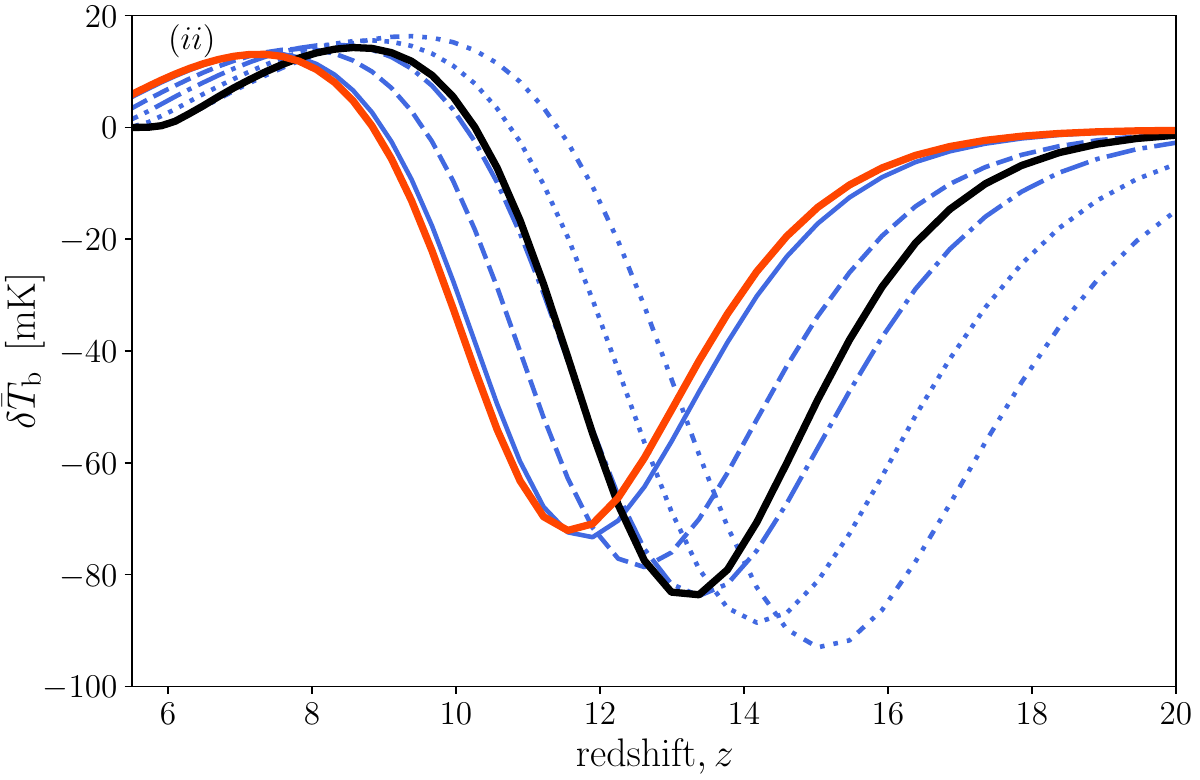}	
	\includegraphics[trim = -0.3cm 0cm 0cm 0cm, scale = 0.7]{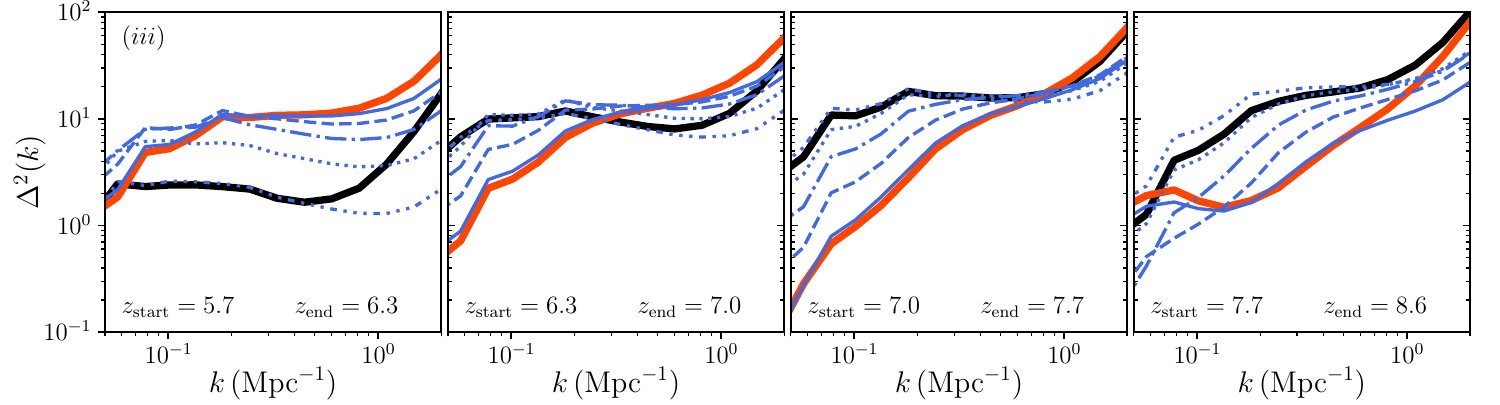}
	\includegraphics[trim = 0cm 0.3cm 0cm 0cm, scale = 0.585]{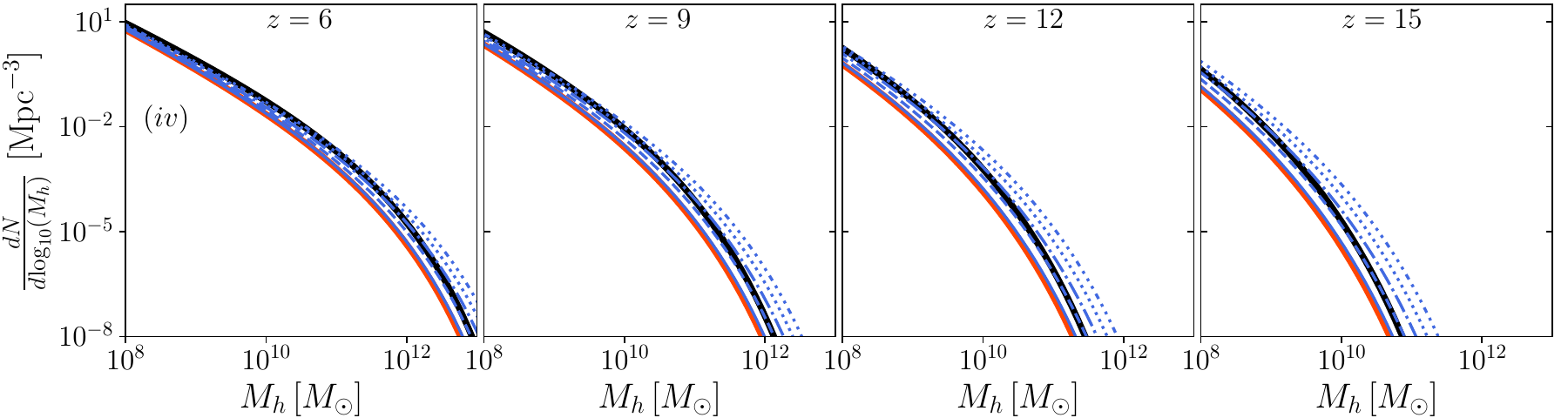}		
    \caption{The impact of changing the HMF constant which controls the exponential cut-off, $d$. We compare the behaviour of this parameter relative to the fiducial astrophysical model with a Sheth-Tormen HMF (black curve) and also a four parameter HMF parameterisation (red curve) with one HMF parameter removed. Panel (\textit{i}) corresponds to the reionisation history, (\textit{ii}) corresponds to mean 21-cm brightness temperature signal, (\textit{iii}) shows the 21-cm PS for a couple of redshifts and (\textit{iv}) the HMF. We consider ${\rm log}_{10}(d) = (-0.45, -0.5, -0.55, -0.6, -0.65)$, demonstrating the sensitivity of the 21-cm signal to this parameter.}
    \label{fig:explore-d}
\end{figure*}

In order to better understand the sensitivity of the model parameters to the exponential term, $d$, and the preference for a decreasing $d$ between the four and five parameter HMF models, we explore the sensitivity of this parameter to the various summary statistics of the 21-cm signal. For this, we consider the same astrophysical and HMF model parameter values as the mock observation with the ST HMF (except with $e=0$ for the four parameter HMF model). In Figure~\ref{fig:explore-d} we consider: (i) the reionisation history, (ii) the global brightness temperature, (iii) the 21-cm PS and (iv) the corresponding HMFs. In all, the black (red) curves correspond to the five (four) parameter HMF model. For this four parameter HMF, we then vary ${\rm log}_{10}(d)$ in increments of -0.05 from -0.45 to -0.65, as highlighted by the light-blue curves.

Quite simply, by removing the fifth parameter ($e=0$) while keeping all others parameter values fixed results in a considerably different behaviour for the 21-cm signal. By decreasing the value of $d$ (i.e. softening the exponential cut-off of the HMF) from the expected value in the four parameter model we can quickly recover statistics of the 21-cm signal consistent with the five parameter model. However, modifications to $d$ alone cannot do all the work, and we further require other shifts in the model parameters in order to match the amplitude and shape of the 21-cm PS. Decreasing $d$ leads to an increasing relative amplitude for the HMF which increases the number of sources (e.g. UV ionising and X-ray photons) causing reionisation and other 21-cm milestones to occur earlier. Therefore, owing to the sensitivity of the HMF to $d$, it is both the most strongly constrained and also most susceptible to being offset from its true value. The behaviour observed here, is consistent with the trends in the biased parameters observed for the four parameter model found above. The lower value of $d$ increases the amplitude in the HMF allowing for the EoR to occur earlier, however in order not to overproduce the too many haloes, we must subsequently shift to a higher $M_{\rm turn}$.

\subsection{Differentiating between HMFs}

As established earlier, fixing the underlying HMF can lead to the biased recovery of the astrophysical parameters. Alternatively, one can attempt to simultaneously constrain the astrophysical parameters and a generalised form for the HMF in order to mitigate the potential biases, at the cost of considerably weakened constraints (factor of $\sim2-4$). Although the marginalised posteriors on the individual HMF parameters are rather broad, the important question is whether or not these translate to marginalised posteriors on the HMF as a whole which are capable of distinguishing between various HMFs in the literature.

Again, to obtain our posteriors on the HMFs given our database of simulated models, we are required to retrain \textsc{Swyft} (equivalent to extracting the SFRD posteriors in Section~\ref{sec:SFRD}). Strictly speaking, we train a new set of networks for each HMF at each redshift. In the upper-right corner of Figure~\ref{fig:GeneralHMF-noLF} we present the marginalised 68th (95th) posteriors for the HMF obtained from the same mock 21-cm PS observation at $z=6, 9, 12$ and 15. The grey (orange) curves correspond to the five (four) parameter model for the HMF respectively. For reference, we additionally overlay the five HMFs considered in our earlier analysis, namely ST (black), PS (red), Watson-$z$ (blue), Angulo (magenta) and Tinker (teal).

Immediately evident from Figure~\ref{fig:GeneralHMF-noLF} is that the marginalised posteriors on the HMF are broader than the difference between the various analytic HMFs considered in this work. Therefore, simultaneously constraining the astrophysical and HMF model parameters from just a 21-cm PS observation with the SKA will not be sufficient to clearly distinguish between the various HMF models available in the literature. However, despite the relatively broad posteriors on the individual HMF model parameters, the marginalised posteriors on the actual HMF are fairly tightly constrained around the input ST HMF. Note, in both cases, we find the recovered HMF posteriors marginally overestimate the true underlying HMF, with the level of discrepancy increasing for higher redshifts. However, this is not unexpected given that we prefer a lower $d$ in both cases, which causes an increase to the amplitude and slope of the HMF for increasing redshift (see the bottom panel of Figure~\ref{fig:explore-d}). Further, as established earlier we also prefer an increase in the overall HMF normalisation, $a$, at the cost of a decreased $f_{\ast}$ and $f_{\rm esc}$ which further drives up the recovered HMF posteriors. Nevertheless, this demonstrates that simultaneously inferring the astrophysical and HMF model parameters from the 21-cm PS alone does a reasonable job at constraining the underlying HMF.

\subsection{Reducing the normalisation parameters}

Earlier, we found that allowing $a$, $f_{\ast,10}$ and $f_{\rm esc,10}$ to be independently constrained in our model resulted in notably large offsets in their recovered values. Therefore now, we repeat our analysis, reducing these three parameters down to two, the product $a\times f_{\ast,10}$ and $f_{\rm esc,10}$. Within \cmfst{} $f_{\ast,10}$ is always multiplicative with the HMF, as it is used to determine star-formation, either in the production of UV or X-ray photons, thus these two quantities will always be correlated. $f_{\rm esc,10}$ on the other hand only becomes important during the EoR, thus it can act somewhat independently. 

\begin{figure}
	\includegraphics[trim = 0.2cm 0.5cm 0.7cm 0.4cm, scale = 0.49]{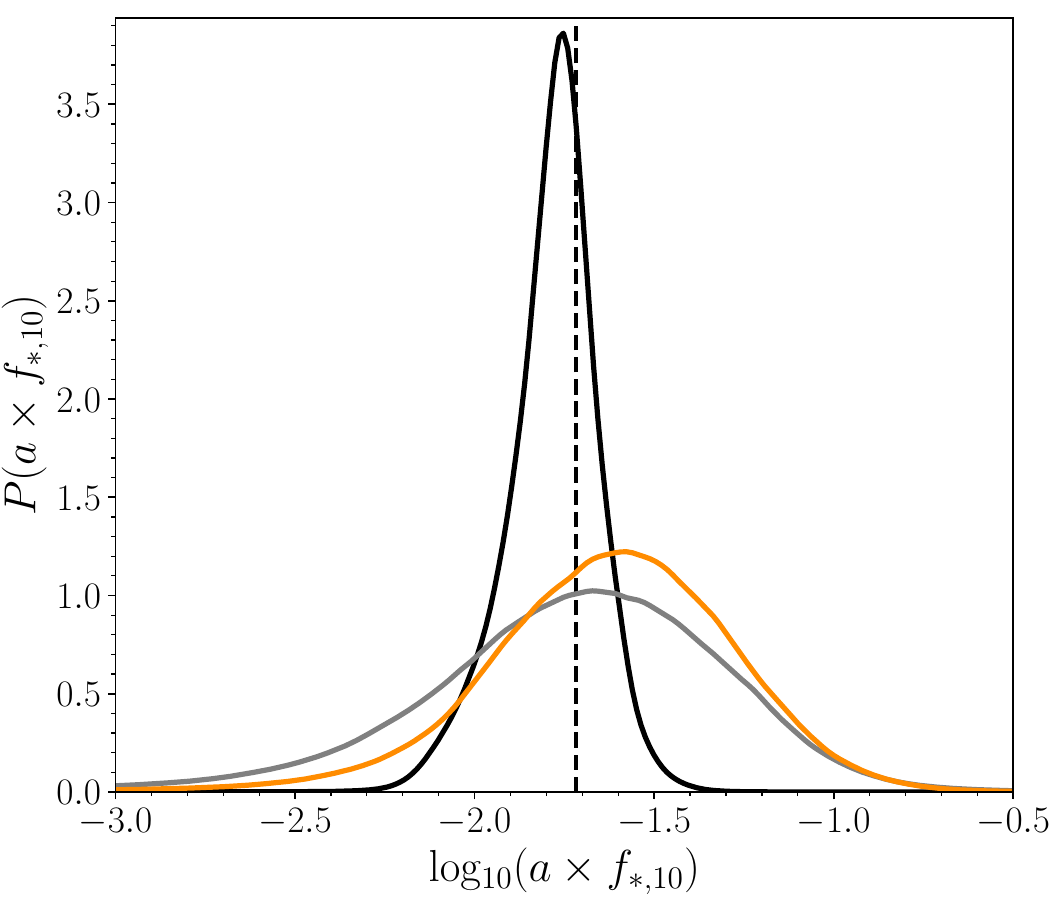}
    \caption{The recovered marginalised 1D posteriors on the joint normalisation, $a\times f_{\ast,10}$, obtained after retraining \textsc{Swyft} for include this normalisation product in our joint astrophysical and general HMF model assuming a mock 1000 hr observation of the 21-cm PS assuming foreground avoidance with the SKA. The grey (orange) curves correspond to our five (four) parameter HMF (see text for further details). The vertical black dashed line denote our fiducial value for this joint normalisation and the black curves are the posteriors from Figure~\ref{fig:21cmPS-noLF} when assuming the correct ST HMF multiplied by the known ST HMF normalisation. Note, we only show the 1D marginalised posteriors for $a\times f_{\ast,10}$ as the 1 and 2D marginalised posteriors for all remaining model parameters should remain mostly unchanged (modulo SBI stochasticity) as those in Figure~\ref{fig:GeneralHMF-noLF} as the underlying training set remains unchanged.}
    \label{fig:GeneralHMF-norm}
\end{figure}

In Figure~\ref{fig:GeneralHMF-norm} we provide the 1D marginalised posteriors for the joint normalisation, $a\times f_{\ast,10}$, for the same mock observation of the 21-cm PS with the SKA after retraining \textsc{Swyft} for our updated joint astrophysical and HMF model. Again, the black curve corresponds to the assuming the correct ST HMF (black curves, Figure~\ref{fig:21cmPS-noLF}) after multiplying the recovered $f_{\ast,10}$ by the known ST HMF normalisation) whereas the black dashed vertical line is the fiducial joint normalisation. The gray (orange) curves correspond to the 5 (4) parameter HMF model. Note, here we only show the 1D marginalised posteriors for $a\times f_{\ast,10}$ as the 1 and 2D marginalised posteriors for all remaining parameters should remain the same as those in Figure~\ref{fig:GeneralHMF-noLF} as the underlying training set remains unchanged. However, in practice owing to inherent stochasticity within the SBI framework, some minor differences are recovered. Nevertheless, in Table~\ref{tab:results-general} we summarise the 68th percent marginalised uncertainties for this full model parameter set with joint normalisation. After considering, $a\times f_{\ast,10}$, we now recover unbiased constraints on this joint normalisation, centred around the fiducial value. This is a notable improvement over the large offsets recovered when considered these parameters individually (Figure~\ref{fig:GeneralHMF-noLF}). Therefore, as expected, considering this normalisation product has improved the visual performance of our overall model.

\subsection{Mock 21-cm PS plus UV LF observation}

\begin{figure*}
	\includegraphics[trim = 0.75cm 0.2cm 0.7cm 0.4cm, scale = 0.89]{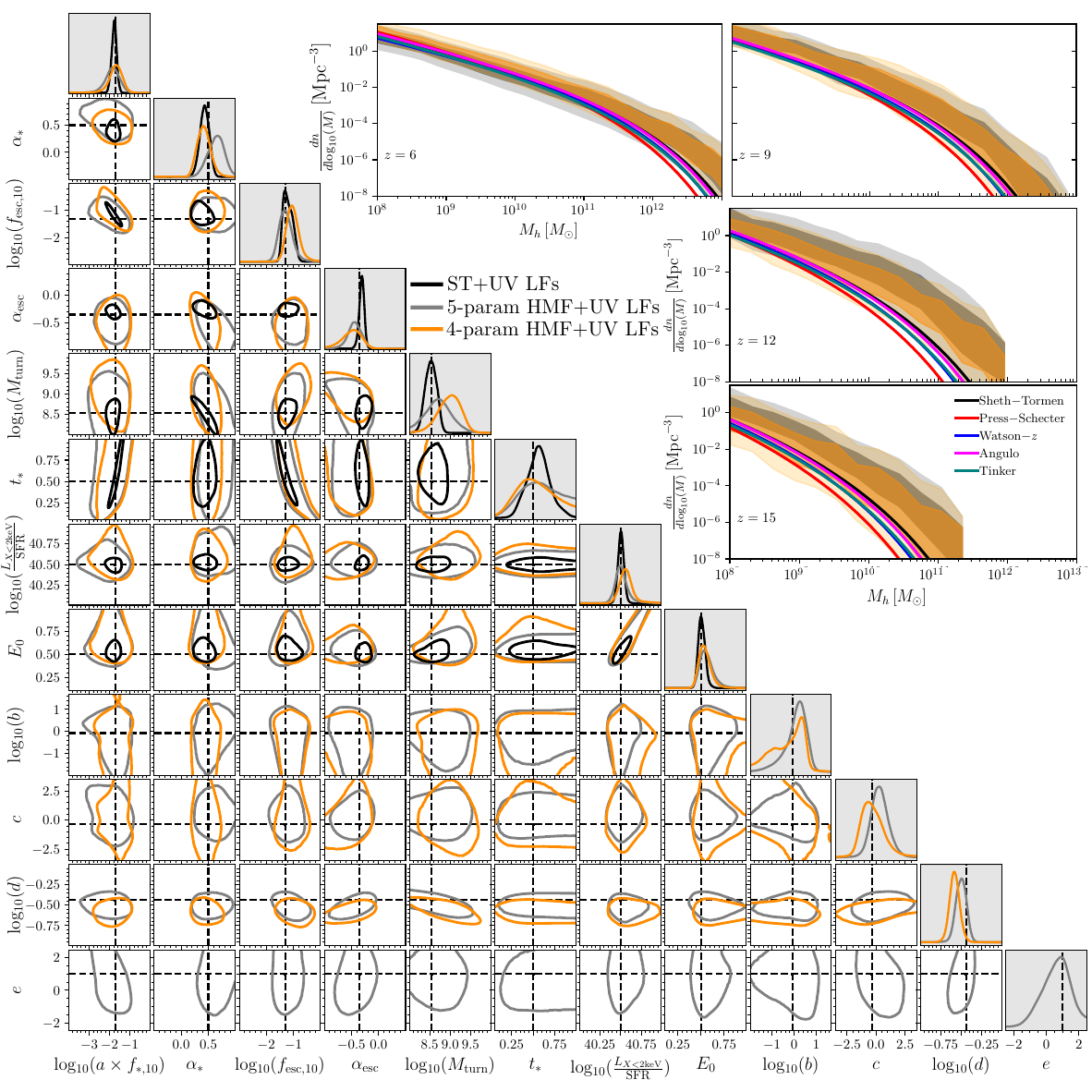}
    \caption{The same as Figure~\ref{fig:GeneralHMF-norm} except also including the UV LFs at $z=6, 7, 8$ and~10 as observational priors.}
    \label{fig:GeneralHMF-wLF}
\end{figure*}

Finally, we consider the impact on the recovered model parameters following the inclusion of UV LF data to our mock observation. In Figure~\ref{fig:GeneralHMF-wLF} we provide the marginalised 1 and 2D posteriors for our mock 21-cm PS observation plus the same UV LFs as considered previously (namely at $z=6,7,8$ and 10). Specifically, we only present the results when considering the normalisation product, $a\times f_{\ast,10}$. However, for completeness we include both this joint normalisation and when keeping $a$ and $f_{\ast,10}$ independent in Table~\ref{tab:results-general}. Finally, in the upper right corner, we also provide the marginalised 68th and 95th percentiles on the resultant HMF following the joint recovery of the astrophysical and HMF model parameters.

Generally speaking, following the inclusion of the UV LFs into our mock observation, the relative improvements in our constraints are more modest compared to a model considering only the astrophysical parameters (i.e. assuming a fixed choice of HMF). However, this is driven by the increased model complexity in trying to simultaneously recover both the astrophysical and HMF parameters. Nevertheless, including the UV LFs improve the overall constraints primarily on $\alpha_{\ast}$ and $a\times f_{\ast,10}$, which subsequently improves the constraints on $\alpha_{\rm esc}$ and $f_{\rm esc,10}$ by breaking the degeneracy. For example, in the absence of the UV LFs, we recovered a relatively larger and more skewed $\alpha_{\ast}$ (0.72) for the five parameter HMF, however using the UV LF data, this almost completely removes this offset with $\alpha_{\ast}$ now $\sim0.55$ compared to the assumed 0.50. 

Both the four and five parameter HMF models still exhibit offsets in ${\rm log}_{10}(d)$ as we have encountered previously. For example, for the five parameter model we recover $-0.49\pm0.06$ which remains borderline consistent with the 68th percentile uncertainties to the fiducial value of $-0.44$, However, the four parameter HMF model is now inconsistent at $\sim3\sigma$, preferring $-0.59\pm0.05$. 

These lower than expected values of ${\rm log}_{10}(d)$ affect our ability to constrain all other remaining model parameters, for example pushing $M_{\rm turn}$ to larger values to compensate for the increasing HMF function amplitude. In turn, this leads to an increasing $f_{\rm esc}$ to allow more UV photons to escape to account for the fact we have fewer lower mass galaxies (increased $M_{\rm turn}$). For the five parameter HMF, owing to the more modest offset this manifests as a lower $a\times f_{\ast,10}$ and also a lower $\alpha_{\rm esc}$ which increases the escape fraction of the lower mass galaxies. On the other hand for the four parameter HMF, we need a significant increase in $f_{{\rm esc},10}$ to compensate for the much larger offset in $M_{\rm turn}$ and $d$. Therefore, removing a model parameter from our generalised HMF (i.e. the parameter $e$) results in a less flexible function form for the HMF leading to stronger biases in the inferred astrophysical parameters. As a result, if one attempts to simultaneously recover the astrophysical and HMF model parameters, the assumed functional form for the HMF must be sufficiently flexible.

Following the inclusion of the UV LFs, focussing on the five parameter model which matches the functional form of the assumed ST HMF, jointly recovering the astrophysical and HMF model parameters results in fractional uncertainties on our astrophysical parameters of between a factor of $\sim2-3\times$ larger than assuming a fixed HMF. Importantly, the recovered marginalised posteriors from simultaneously fitting the HMF jointly with the astrophysical parameters encapsulate the posteriors when fixing the HMF to the correct ST HMF. Further, they recover the same relative degeneracies as evidenced by the similar posterior contours. These more broadened posteriors additionally account for the relative biases in the astrophysical parameters when we previously incorrectly assumed the form of the HMF. That is, the biased posteriors recovered in Figure~\ref{fig:21cmPS-wLF} are contained within these broadened posteriors. Except for the PS HMF, which clearly prefers lower than allowed values of $M_{\rm turn}$. Therefore, by relaxing the assumption of choosing a fixed HMF we are able to obtain more conservative, unbiased constraints on our astrophysical parameters.

In principle, one could also consider other observational constraints on our joint model parameters, which could further improve our overall constraining power limiting the impact of increasing the model complexity to jointly recover the HMF. Nevertheless, these relative increases in the marginalised uncertainties, when jointly recovering the HMF, are not prohibitive when it comes to interpreting real-world data. In fact, in effectively being agnostic to the underlying HMF, the precision of these recovered constraints are more conservative with respect to our ignorance of our modelling uncertainties. However, further work is required to determine the preference for a lower than expected value for the exponential cut-off parameter, $d$ and ways to mitigate/eliminate it.

Focussing on the recovered HMF posteriors, we see an almost factor of three reduction in the posterior width at $z=6$ with these posteriors tightly bounding the various HMFs in the literature. However, the relative improvements decrease for increasing redshift, highlighting that the vast majority of the constraining power comes from the UV LFs at $z=6$. We do observe a more modest shrinking of the posteriors at $z=9$, where we still have observational data (at $z=10$), but no relative improvement at $z=12$ or $z=15$. Therefore, to be able to distinguish between the various HMFs in the literature, we would require improved UV LF constraints spanning out to higher redshifts. Already, recent JWST results are beginning to place limits on the UV LFs extending to redshifts higher than considered here \citep[e.g.][]{Naidu:2022, Donnan:2022, Castellano:2022, Atek:2022, Harikane:2022a, Labbe:2022,Bouwens:2023,Willott:2023}. Further, relaxing our assumption of only considering UV galaxies at $M_{\rm UV} < -20$ allowing us to access the bright end galaxies might further help improve our ability to constrain the underlying HMF from our mock 21-cm PS observation with the SKA by constraining the higher mass end. We shall look into the improvements available when considering both of these in future work.

\section{Conclusions} \label{sec:conclusion}

We are rapidly approaching the first detection of the 21-cm signal during reionisation. To gain insights into the properties of the first generation of galaxies responsible for driving reionisation we are required to develop Bayesian inference pipelines to interpret the observation data. Crucially, when developing these inference frameworks we are required to make various assumptions and simplifications in our theoretical models in order to make them computationally efficient enough to be used for inference. However, it is crucial that we understand the consequences of these choices. One such assumption that is typically adopted is that of a fixed halo-mass function (HMF). Which implies that our inferred model parameter constraints are tied to our assumed choice of HMF.

Within this work, we explored the consequences of incorrectly assuming the HMF on the inferred astrophysical parameters assuming a mock observation of the 21-cm power spectrum (PS). Specifically, we consider a 1000 hr observation of the 21-cm PS assuming foreground avoidance with the forthcoming SKA. Throughout we modelled the 21-cm signal using \cmfst{} \citep{Mesinger:2007p122,Mesinger:2011p1123,Murray:2020} adopting the flexible UV galaxy model parameterisation of \citet{Park:2019}. This work extends on earlier work by \citet{Mirocha:2021} who explored the impact of assuming a HMF along with the stellar population synthesis model on the global 21-cm signal. However, this used a simplified model for simulating the 21-cm signal, whereas here we perform full 3D reionisation simulations in order to study the 21-cm PS. To recover our inferred parameter constraints we performed simulation based inference using the publicly available  software, \textsc{Swyft} \citep{Miller:2021} which performs marginal neutral ratio estimation (MNRE) to learn the likelihood-to-evidence ratio of our simulated data to obtain the marginalised posteriors on our astrophysical parameters.

Adopting the ST HMF \citep{Sheth:2001} as our fiducial model HMF we then recovered our model astrophysical parameters assuming five difference choices for the underlying HMF in order to encapsulate the breadth of different models in the literature. In addition to the ST HMF, we also considered the PS \citep{Press:1974}, Angulo \citep{Angulo:2012}, Tinker \citep{Tinker:2010} and the redshift evolving Watson HMF \citep{Watson:2013}. When considering just the 21-cm PS alone, we encountered strong biases of up to $\sim3-4\sigma$ on our recovered astrophysical parameters, most notably in the mass dependence of the escape fraction, $\alpha_{\rm esc}$ and $M_{\rm turn}$ the characteristic turn-over mass describing the typical masses of star-forming galaxies.

The source of these biases are driven by the underlying model seeking to match the star-formation rate density over the entire cosmic history. The differences in the HMF lead to the cosmic 21-cm milestones (reionisation, X-ray heating or \lya{} coupling) happening earlier or later depending on if they over or under predict the number density of sources. Thus, the biases in astrophysical parameters compensate for the HMF differences to ensure the cosmic star-formation rate density correctly matches that of the mock observation.

Next, we folded in observed UV LFs with our 21-cm PS observation in order to improve our constraints on our model parameters. Primarily, the inclusion of the UV LFs constrains the stellar content of the UV galaxies strongly constraining $f_{\ast}$, both through its normalisation and mass dependence. This in turn breaks the corresponding degeneracy with the escape fraction, $f_{\rm esc}$. In doing so, this restricts the resultant posterior volume available to match the true star-formation rate density and compensate for the incorrect HMF choice. Primarily, this drives biases in $M_{\rm turn}$ and $\alpha_{\rm esc}$. Although the relative amplitude of these biases are reduced compared to just the 21-cm PS, the reduced marginalised uncertainties still lead to significant biases of $\sim3\sigma$ for both parameters. 

Given that incorrectly assuming the HMF can lead to large biases in our astrophysical parameters, next we explored whether we could jointly recover the astrophysical parameters along with the underlying HMF from the same mock 21-cm PS observation with the SKA. To do so, we constructed a simple five parameter HMF model and simultaneously recovered these parameters in addition to our eight astrophysical parameters. Further, we also considered the case where we removed one of these five HMF model parameters, in order to explore the consequences of jointly fitting a HMF that is inflexible to the true underlying ST HMF.

By constraining this extended model with a generalised form for the HMF, our astrophysical parameters are recovered with a factor of $\sim2-4$ larger marginalised uncertainties, owing to the increased model complexity. However, in doing so our constraints are now unbiased by our choice of HMF and are thus more conservative. Note, we still recovered small offsets in some of our astrophysical parameters, namely $\alpha_{\ast}$ and $\alpha_{\rm esc}$, but this is driven by a corresponding offset in the opposite direction for one of the HMF model parameters, $d$ which controls the exponential cut-off. In the case of our four parameter model, which is designed to be inflexible, we recovered considerably stronger skews in $M_{\rm turn}$ and $d$. Therefore, by assuming a generalised HMF incapable of matching the observation data, one will cause significantly larger shifts in the inferred model parameters (both astrophysical and HMF).

Finally, following the inclusion of UV LFs into our joint HMF plus astrophysical parameter model, we still find $\sim2-3\times$ larger marginalised uncertainties on our astrophysical parameters. However, importantly, we recover all input astrophysical parameters to within our 68th percentile uncertainties. Therefore, by jointly recovering the astrophysical and HMF model parameters we can recover unbiased constraints on the input parameters. Although the relative uncertainties are notably larger when comparing to assuming a fixed HMF, they are conservative and more importantly agnostic to the choice of HMF. Further, we also recovered the marginalised posteriors on the underlying HMF and although they were relatively tightly constrained, they were broader than the variation between the various HMFs considered earlier in this work and thus we are unable to perform model selection to distinguish the input (ST) HMF.

In future, we could expand on this work by utilising the Bayesian evidence to perform model selection to: (i) attempt to rule out certain incorrectly assumed choices of the HMF or (ii) to determine the necessary complexity of a generic HMF beyond the five parameter model considered within this work. Alternatively, rather than jointly sampling the astrophysical and HMF parameters, if we are only interested in the astrophysical parameters we could consider performing Bayesian model averaging of the extracted posteriors obtained from a broad range of HMF models.

\section*{Acknowledgements}

Parts of this research were supported by the Australian Research Council Centre of Excellence for All Sky Astrophysics in 3 Dimensions (ASTRO 3D), through project number CE170100013. J.M. was supported by an appointment to the NASA Postdoctoral Program at the Jet Propulsion Laboratory/California Institute of Technology, administered by Oak Ridge Associated Universities under contract with NASA. Y.S.T. acknowledges financial support from the Australian Research Council through DECRA Fellowship DE220101520. A.M. acknowledges support from the Ministry of Universities and Research (MUR) through the PRIN project ”Optimal inference from radio images of the epoch of reionization” as well as the PNRR project ”Centro Nazionale di Ricerca in High Performance Computing, Big Data e Quantum Computing”.

\section*{Data Availability}

The data underlying this article will be shared on reasonable request to the corresponding author.


\bibliographystyle{mnras}
\bibliography{Papers} 



\appendix

\section{Assessing Network Coverage} \label{sec:coverage}

\begin{figure*}
	\includegraphics[trim = 0cm 0.3cm 0cm 0.3cm, scale = 0.95]{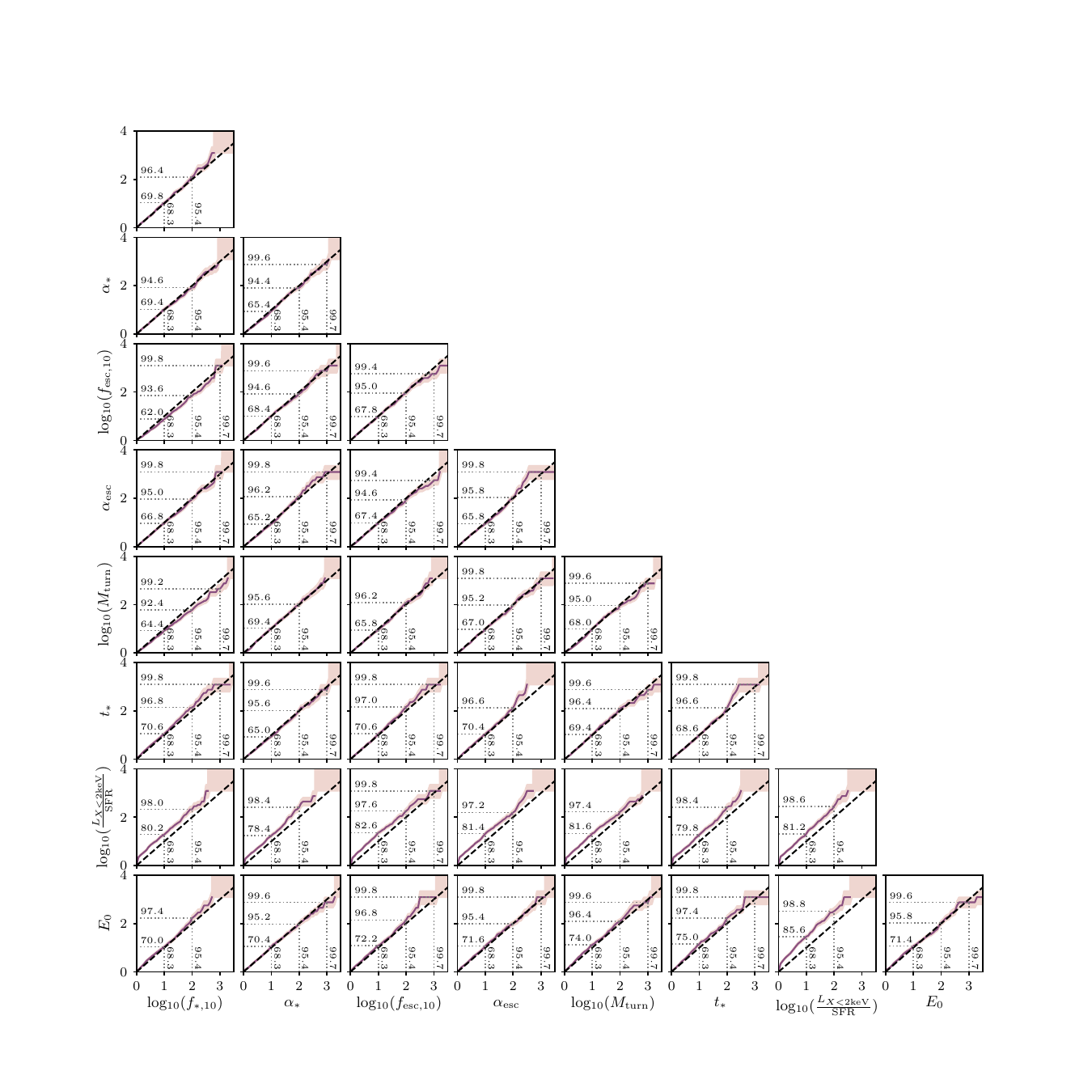}
    \caption{The empirical expected coverage probability of our trained MNRE network with \textsc{Swft} (vertical axis) as a function of the confidence level (horizontal axis). The purple line corresponds to the coverage of our network, with perfect coverage denoted by the diagonal black dashed line, while dotted lines denote the 68th, 95th and 99.7th percentiles while the shaded region corresponds to the Jeffrey's interval (see text for further details).}
    \label{fig:coverage}
\end{figure*}

One of the significant advantages of SBI like MNRE is that once the network is trained they provide rapid evaluation of the posteriors for any given input data. That is, we can perform inference for a large number of input models (not used in the training of the original network) sampling our prior volume and determine the frequency with which each model falls within the predicted posteriors determined by our trained network. Ultimately, this enables us to quantify the coverage of the network \citep[e.g.][]{Cole:2022}, which is a considerably more robust concept than the typical convergence criteria used by traditional MCMC approaches \citep[e.g.][]{Betancourt:2019,Roy:2020}.

First, we define the function, $\Theta_{\hat{p}(\hat{\boldsymbol{\theta}}|\boldsymbol{x}_{i})}(1-\alpha)$, which gives the $(1-\alpha)$ highest probability density region (HPDR) for our estimated posterior, $\hat{p}(\hat{\boldsymbol{\theta}}|\boldsymbol{x}_{i})$, given the input mock data, $\boldsymbol{x}_{i}$, and known input parameters, $\boldsymbol{\theta}^{\ast}_{i}$. For example, a 95 per cent HPDR would have an error rate of $\alpha=0.05$. Then, for an independent set of $n$ randomly drawn models, $(\boldsymbol{x}_{i},\boldsymbol{\theta}^{\ast}_{i})$ we determine:
\begin{eqnarray}
1 - \hat{\alpha} = \frac{1}{n}\sum^{n}_{i=1} \mathds{1} \left[\boldsymbol{\theta}^{\ast}_{i} \in \Theta_{\hat{p}(\hat{\boldsymbol{\theta}}|\boldsymbol{x}_{i})}(1-\alpha) \right],
\end{eqnarray}
the actual error rate of the HPDR given our estimated posterior.

The quantities $\alpha$ ($\hat{\alpha}$) can be re-parameterised in terms of a new variable, $z$, which corresponds to the $1-\alpha/2$ ($1-\hat{\alpha}/2$) quantile of the standard normal distribution. In this way, the 1, 2, 3$\sigma$ regions correspond to $z=1,2,3$ with $1-\alpha = 0.6827, 0.9545, 0.9997$. The uncertainties on $\hat{\alpha}$ are determined using the Jeffreys interval \citep{Cole:2022}\footnote{This interval is the 68.27 per cent central interval of a Beta distribution defined by the parameters ($n-k+1/2,k+1/2$) where $n$ is the total number of samples from the joint model and $k$ is the number of times the HPDR predicted by the network does not contain the true astrophysical parameters.}. In Figure~\ref{fig:coverage} we present the empirical expected coverage probability of our trained network for our mock 21-cm observation assuming an ST HMF as a function of confidence levels for all 1D and 2D marginalised posteriors. The goal is to achieve a network coverage (purple line) that follows the black dashed curve (perfect coverage). Deviations of the resultant network coverage above the dashed black line implies our network is conservative with respect to the recovered posteriors whereas deviations below are considered overconfident. For our particular setup, we recover excellent network coverage, with near perfect performance for most 2D parameter combinations.

\begin{figure*}
	\includegraphics[trim = 0.3cm 0.3cm 0cm 0cm, scale = 0.89]{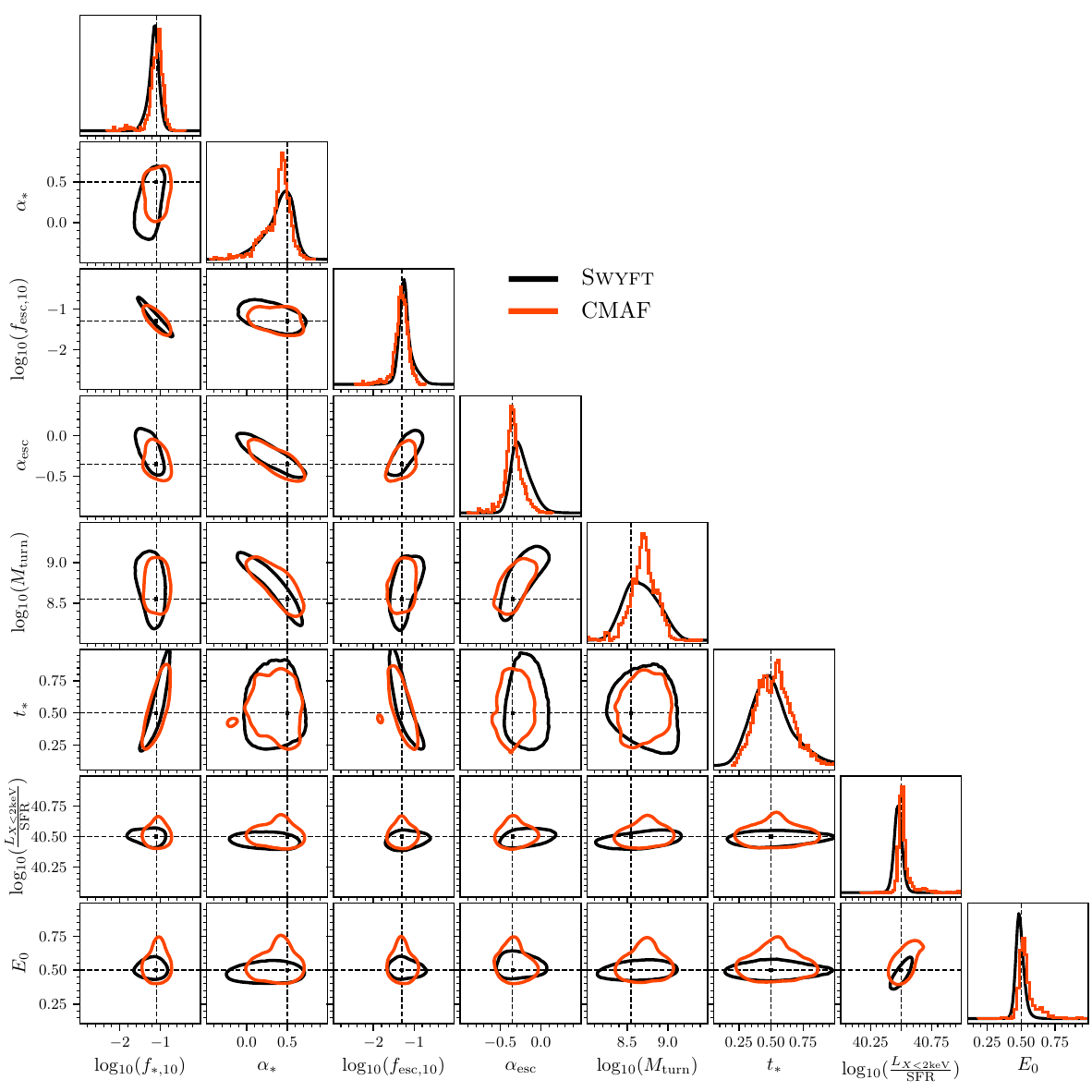}
    \caption{Comparison of the 1 and 2D marginalised posteriors of our fiducial astrophysical model assuming our mock 21-cm PS observation with the SKA obtained using two different SBI approaches: \textsc{Swyft} (black contours) and conditional masked-autoregressive flow (red contours).}
    \label{fig:CMAF}
\end{figure*}

\section{Comparison of SBI approaches} \label{sec:CMAF}

During the completion of this work we simultaneously employed MNRE with \textsc{Swyft} to learn the likelihood-to-evidence ratio while also applying the conditional masked-autoregressive flow (CMAF) model from \citet{Prelogovic:2023} to explicitly learn the likelihood. CMAF is a density estimator which performs a series of linear transformations of normal random variables to estimate a target distribution (see e.g. \citealt{Papamakarios:2021} for a recent review). Using the publicly available code, \textsc{21cmLikelihoods} \footnote{https://github.com/dprelogo/21cmLikelihoods} we performed SBI on the same simulated dataset of models as outlined in Section~\ref{sec:sim_data} to learn our likelihood. From this, we then use \textsc{EMCEE} \citep{ForemanMackey:2013p823} assuming the same priors to obtain our posteriors given the same mock 21-cm PS observation.

In Figure~\ref{fig:CMAF} we provide the recovered marginalised 1 and 2D marginalised posteriors from \textsc{Swyft} (black curves) and after learning the likelihood using CMAF (red curves). Below the diagonal, we provide the 95th percentile joint marginalised posteriors. Generally speaking we recover consistent results between the two different SBI approaches, with \textsc{Swyft} typically producing slightly broader posteriors. Given the vastly different methodologies employed by both methods to achieve similar results, this yields additional confidence in both approaches. In the end, we found \textsc{Swyft} to be more computationally efficient and also more stable to train and thus adopted it for the remainder of this work. It is likely that the CMAF approach requires additional ensemble learning or active learning as explored in \citep{Alsing:2019} to boost its stability.

\section{Modifying UV LFs for SBI} \label{sec:LF}

SBI requires a stochastic simulator of the modelled data given a set of input parameters. However, in \cmfst{} our UV LFs are derived from analytic functions as outlined in Equation~\ref{eq:UVLF}. Therefore, to translate the observational uncertainties from real-world observations into our simulated UV LFs we simply add random noise to each $M_{\rm UV}$ bin (see Section~\ref{sec:addUVLFs}). Here, we verify this approximation by using \textsc{Swyft} to obtain our marginalised posteriors on our astrophysical model parameters using only the UV LF data in our training set and compare the results to an MCMC where we produce the UV LFs from \cmfst{} on-the-fly using \textsc{EMCEE} and assume a simple analytic likelihood (i.e. $\chi^{2}$, \citealt{Park:2019}).

\begin{figure}
	\includegraphics[trim = 0.3cm 0.5cm 0cm 0cm, scale = 0.68]{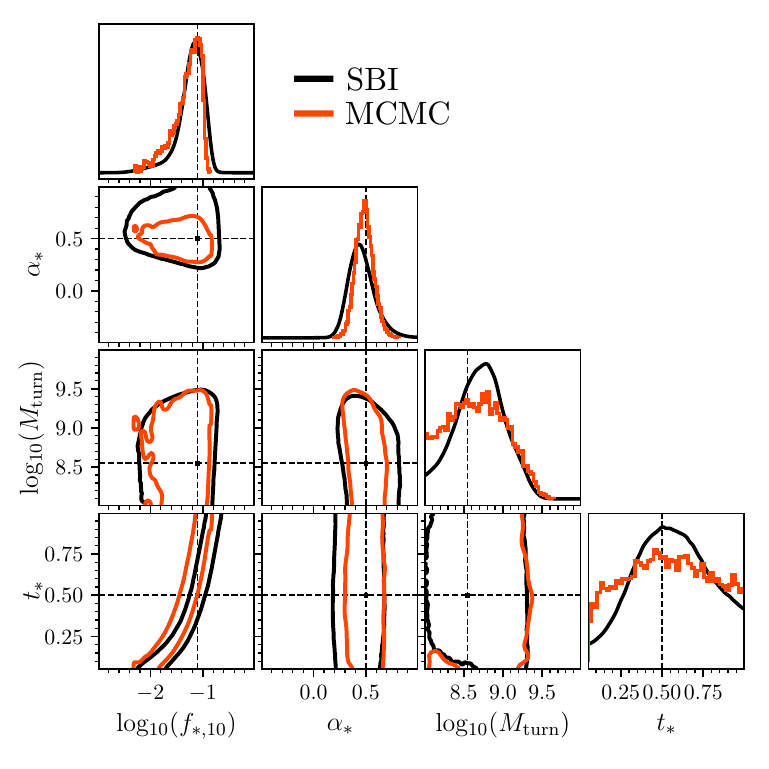}
    \caption{Comparison of our marginalised posteriors obtained from only considering UV LFs. The black curves correspond to performing SBI with \textsc{Swyft} after modifying the simulated data to directly include observational uncertainties and the red curves correspond to a direct MCMC assuming an analytic form for the likelihood.}
    \label{fig:LFcomp}
\end{figure}

In Figure~\ref{fig:LFcomp} we compare the 1 and 2D marginalised posteriors (95th percentiles) using \textsc{Swyft} (black curves) and our direct MCMC approach (red curves). The consistency of the posteriors following the two approaches provides confidence that our procedure for adding noise to the analytic UV LFs in order to be able to include them within our SBI framework is sufficiently accurate. Note, we do observe slightly tighter 1D marginalised posteriors for $M_{\rm turn}$ and $t_{\ast}$ with \textsc{Swyft} implying improved constraining power over direct MCMC, however, the differences are fairly minor. Nevertheless, these differences likely arise because of the implicit likelihood that the MNRE approach of \textsc{Swyft} is extracting, which demonstrates the importance of going beyond the simple $\chi^{2}$ likelihood form of traditional MCMCs. The differences between the likelihoods are non-trivial, but in applying MNRE we always reduce corresponding biases resulting in more precise posteriors.


\bsp	
\label{lastpage}
\end{document}